\documentstyle[12pt,epsfig,epsf,axodraw]{article}

\renewcommand{\thefootnote}{\fnsymbol{footnote}}

\newcommand{\ra}{\rightarrow}
\newcommand{\ee}{e^+e^-}
\newcommand{\Dep}{\tilde\Delta^{++}}
\newcommand{\dep}{\tilde\delta^{++}}
\newcommand{\Dem}{\tilde\Delta^{--}}
\newcommand{\dem}{\tilde\delta^{--}}
\newcommand{\x}{\tilde{\chi}^{0}}
\newcommand{\se}{\tilde{e}}
\newcommand{\sle}{\tilde{\l}}

\newcommand{\mrm}[1]{\mbox{\rm #1}}
\newcommand{\beq}{\begin{equation}}
\newcommand{\eeq}{\end{equation}}
\newcommand{\nn}{\nonumber}
\newcommand{\bea}{\begin{eqnarray}}
\newcommand{\eea}{\end{eqnarray}}
\newcommand{\eq}[1]{eq.(\ref{#1})}
\newcommand{\rfn}[1]{(\ref{#1})}

\newcommand{\ea}{{\it et al.}}

\newcommand{\ie}{{\it i.e.}}

\newcommand{\np}[1]{{ Nucl. Phys. }{\bf #1}}
\newcommand{\plet}[1]{{ Phys. Lett. }{\bf #1}}
\newcommand{\pr}[1]{{ Phys. Rev. }{\bf #1}}
\newcommand{\prlet}[1]{{ Phys. Rev. Lett. }{\bf #1}}
\newcommand{\zp}[1]{{ Z. Phys. }{\bf #1}}
\newcommand{\prep}[1]{{ Phys. Rep. }{\bf #1}}
  
\newcommand{\ijmp}[1]{{ Int. J. Mod. Phys. }{\bf #1}}

\def\lsim{\mathrel{\vcenter{\hbox{$<$}\nointerlineskip\hbox{$\sim$}}}}
\def\gsim{\mathrel{\vcenter{\hbox{$>$}\nointerlineskip\hbox{$\sim$}}}}

\begin{document}

\newpage
\setcounter{page}{0}

\begin{titlepage}
\begin{flushright}
\hfill{DESY 98-136}\\
\hfill{\today}
\end{flushright}
\vspace*{1.0cm}

\begin{center}
{\large \bf 
Doubly Charged Higgsino Pair Production and Decays \\
in \boldmath ${\bf e^+e^-}$ Collisions}
\end{center}
\vskip 1.cm
\begin{center}

{\sc M. Raidal}\footnote{A.v.Humboldt Fellow} and     
{\sc P.M. Zerwas}

\vskip 0.8cm

\begin{small} 
 DESY, Deutsches Elektronen-Synchrotron, D--22603 Hamburg, Germany 
\end{small}
\end{center}

\vskip 2cm

\setcounter{footnote}{0}
\begin{abstract}
In supersymmetric grand unified theories,
light higgsino multiplets generally exist in addition to the familiar 
chargino/neutralino multiplets of the minimal supersymmetric 
extension of the Standard Model. The new multiplets may include
doubly charged states  $\tilde\Delta^{\pm\pm}$ and    
$\tilde\delta^{\pm\pm}$. We study the properties and the 
production channels of these novel higgsinos in $e^+e^-$ and
$\gamma\gamma$ collisions, and investigate how their properties 
can be analyzed experimentally.
\end{abstract}

\end{titlepage}

\newpage
\renewcommand{\thefootnote}{\alph{footnote}}

%%%%%%%%%%%%%%%%%%%%%%%%%%%%%%%%%%%%%%%%%%%%%%%%%%
\subsection*{1. Synopsis}
%%%%%%%%%%%%%%%%%%%%%%%%%%%%%%%%%%%%%%%%%%%%%%%%%%

While the Standard Model (SM) has been extremely successful in 
interpreting  nearly all experimental observations in the past
three decades, there is  increasing {\it experimental} evidence that the model
should be embedded in a more comprehensive theory.
The  deficit of solar  and atmospheric 
neutrino fluxes
and  indications for the existence of hot and
cold components of dark matter  point clearly to 
directions of physics beyond the Standard Model \cite{neutrino}.

Strongly motivated by the observations of non--zero neutrino masses, 
the embedding of the SM in a left--right (LR) symmetric
 grand unified 
theory (GUT) like $SO(10)$ \cite{lr,so10} or $E_6$  \cite{e6}
is a most attractive direction.
In this approach, the chiral fermion fields of one generation
are grouped together in a single multiplet of the 
fundamental representation, including the right--handed neutrino
component.

The hierarchy problem, related to light fundamental scalars in the context
of very high GUT scales, is partly solved, in a natural way,
by extending the model to a supersymmetric (SUSY) theory.
The non--trivial vacuum  structure and the breaking of supersymmetry
leads to interesting new phenomena. 
In SUSY GUT-s with  intermediate left--right  symmetry, novel
light superfields can be present despite the very
high scale of the left--right symmetry breaking \cite{sen1,sen2,sen3}. 
The resulting low--energy theory is the $R$--parity conserving
minimal supersymmetric standard model (MSSM), supplemented
by light massive neutrinos which can be generated by the see--saw
mechanism \cite{seesaw},
and light remnants of the Higgs supermultiplets. If the scale for 
left--right symmetry breaking is chosen such as to generate the
right order of neutrino masses, the new light states should
have masses in the range of $\sim$ 100 GeV.

In this study we focus on one of the central predictions of 
this type of SUSY models in the light fermionic higgsino sector: 
light doubly charged $SU(2)_L\times U(1)_Y$ 
triplet components  $\tilde\Delta^{++}$ and  
singlets $\tilde\delta^{++}$. 
We construct the effective low--energy model incorporating
these new fields and define their interactions. 
The influence of the new particles on the unification of
couplings is studied in the context of SUSY SO(10). Subsequently
we work out the phenomenology of these particles at future 
$e^+e^-$ linear colliders, extending earlier work in 
Refs.\cite{higgsino1,higgsino2}:
\begin{eqnarray} 
\label{DD}
\ee&\ra&\Dep\Dem \\
\label{dd}
\ee&\ra&\dep\dem
\end{eqnarray}
We will discuss the production of the doubly charged higgsinos and 
their decay modes\footnote{The phenomenology of the 
doubly charged Higgs bosons has 
previously been studied  extensively in Refs.\cite{higgs++}.}.
Final--state correlations among the decay 
products, rooted in spin--spin correlations, can be
exploited to measure the fundamental couplings of these particles
\cite{R7A}.
$\gamma\gamma$ collisions which are particularly suited for the 
production of doubly charged particles will also be briefly commented.

The outline of the paper is as follows. After describing the general 
physics base in Section 2, the production of higgsinos will
be presented in Section 3, followed by a discussion of the decay
modes in the subsequent section. In Section 5, angular correlations
will be exploited to determine the higgsino couplings.

%%%%%%%%%%%%%%%%%%%%%%%%%%%%%%%%%%%%%%%%%%%%%%%%%%
\subsection*{2. Effective Low--Energy  Theory}
\label{sec:2}
%%%%%%%%%%%%%%%%%%%%%%%%%%%%%%%%%%%%%%%%%%%%%%%%%%

Grand unified theories which incorporate left--right  symmetries,
include the groups $SO(10)$ and $E_6$. If broken down
to the Standard Model gauge group $SU(2)_L \times U(1)_Y$, effective
theories based on intermediate
$SU(2)_L\times SU(2)_R\times U(1)_{B-L}$ or other
symmetries may be realized at a scale $M_R$.
Embedding such models in supersymmetric theories 
can solve many of the problems of the MSSM: strong and weak
CP problems; the conservation of $R$ parity \cite{moha1}.
These theories can also accommodate small neutrino masses 
through the see--saw mechanism in a natural way.

Several theoretical possibilities exist
in SUSY LR models \cite{susylr}
 which lead, after
symmetry breaking, to vacua  conserving the electric charge. 
(i) Either the LR breaking scale must
be low, $M_R \lsim M_{SUSY}$, and $R$--parity must
be spontaneously broken at the same time \cite{moha2};
or (ii) $B-L$ neutral triplets must be added; or 
(iii) non--renormalizable interactions must be introduced \cite{sen1,sen2}.
The phenomenomena emerging from  (i) 
have been studied in Ref.\cite{kati}. 
The scenario (ii) may lead to
new light singly charged higgsinos.     
In this paper, we concentrate specifically on phenomena following 
from the third solution which involves doubly charged spin--1/2 particles. 

This solution implies that the right--handed symmetry breaking scale
is very high, $M_R \gsim {\cal O} (10^{10} - 10^{11}$ GeV).
At such a scale effective higher-order
operators originating from  Planck scale physics may start playing a role. 
Since the scale $M_R$ sets the natural mass scale of the new particles,
one may na\"{\i}vely guess that all the new particles decouple
from the low--mass spectrum. However, as a result of the 
vacuum structure, the decoupling is not complete in 
supersymmetric theories. (This was already noticed quite early in  
 Ref.\cite{earlyso10}). 
If the supersymmetry is unbroken,
the symmetry leads to an ensemble of degenerate vacua
corresponding to flat directions. The excitations 
associated with these flat directions are massless
particles. If SUSY is broken, the $D$--flat directions
are lifted and the theory picks one of the vacua;
if $R$--parity is conserved and $M_R$ is sufficiently
high \cite{beall}, the vacuum conserves the electric charge \cite{sen1,sen2}.
The previously massless excitations are transformed to states with 
masses of order $m \sim M^2_R / M_{PL}$, with $M_{PL}$ being
the Planck scale. Besides the light neutrinos,
the effective low--energy theory will include these light remnants
in addition to the MSSM particle spectrum.

In this scenario the effective low--energy theory is defined
by the MSSM with exact $R$--parity, supplemented by two left--handed
triplet  superfields  $\Delta$ and $\bar\Delta$, and two right-handed
singlet superfields $\delta$ and $\bar\delta$ with opposite $U(1)$ 
quantum numbers such as to cancel chiral anomalies. They
are assigned the $SU(2)_L \times U(1)_Y$ quantum numbers  
\bea
\label{fields}
\Delta=\left( \begin{array}{cc}
\Delta^+/\sqrt{2} & \Delta^{++} \\
\Delta^0 & -\Delta^+/\sqrt{2}
\end{array}\right) = (3,2)  
& & 
\bar\Delta=\left( \begin{array}{cc}
\bar\Delta^-/\sqrt{2} & \bar\Delta^{0} \\
\bar\Delta^{--} & -\bar\Delta^-/\sqrt{2}
\end{array}\right) = (3,-2)  
\eea
and
\bea
\label{fields2}
\delta=\delta^{--}=(1,-4)  \;\;\;\;\;\;\;\;\;\;\;\;\;{}^{}
& &
\bar\delta=\bar\delta^{++}=(1,4) 
\eea
These fields are the light remnants of the Higgs supermultiplets in
the underlying GUT theory belonging to (3,1) and (1,3) 
representation\footnote{For the $SU(3)_c\times SU(2)_L
\times SU(2)_R \times U(1)_{B-L}$ intermediate theory, the complete
set of quantum numbers read $(1,3,1,\pm 2)$ for $\Delta/\bar\Delta$
and $(1,1,3,\mp 2)$ for $\delta/\bar\delta$, respectively.}
of the intermediate $SU(2)_L\times SU(2)_R$ subgroup, respectively. 
The superpotential, apart from the non-renormalizable terms, may be written as 
\bea
W=W_{MSSM}+W_T+W_S
\label{spot1}
\eea
where $W_{MSSM}$ is the superpotential of the MSSM \cite{mssm}
and the new terms $W_T$ and $W_S$ describing the triplet and singlet
superfield interactions, respectively,  are given by 
\bea
W_T & =& M_{\Delta} \mrm{Tr} \Delta\bar\Delta + i f_\Delta 
L^T\tau_2\Delta L\,,
\nn \\
W_S & =& M_{\delta} \delta^{--}\bar\delta^{++} +  
f_{\delta} l^cl^c\delta^{--}
\label{spot}
\eea 
 The doublet of left-handed leptons is denoted by $L$ and  the
singlet of the right-handed lepton by $l^c$
($f_\delta=f_\Delta$ for strict LR symmetry at the relevant scale). Because
 of  stringent constraints
on lepton--flavor violation from the processes $l_1\to l_2\gamma,$
$l_1\to 3l_2$ and $\mu-e$ conversion in nuclei, the couplings $f$
are diagonal to a high degree of accuracy 
 in family  space \cite{santamar}. 
The experimental
bound from muonium-antimuonium conversion implies the constraint
$f_{ee}f_{\mu\mu}\lsim 1.2\cdot 10^{-3}$ for the mass $M_{\Delta}=100$ GeV
while  no constraints can be set on the coupling $f_{\tau\tau}.$
This implies that light doubly charged particles with masses around 100
GeV
cannot decay to electrons and muons at the same time.  \\

%%%%%%%%Unification%%%%%%%%%%%%%%%

Including new light supermultiplets (\ref{fields},\ref{fields2}) 
in the low--energy particle spectrum will
influence the running of  couplings and may dangerously spoil the
unification of $\alpha_1=3/5\alpha_Y,$ 
$\alpha_2$ and $\alpha_3=\alpha_s$ at the GUT scale \cite{unif},
eventually jeopardizing the unification of the couplings,
or worse, driving the unification point to a dangerously low scale for
proton decay. 
 The evolution of  couplings 
to one--loop order is described by the solutions of the
renormalization group equations,
\bea
\frac{1}{\alpha_i(M_X)}=\frac{1}{\alpha_i(\mu)}+\frac{b_i}{2\pi} 
\ln\left(\frac{\mu}{M_X}\right)
\eea
where the beta functions $b_i$ depend on the particle
content of the theory. 

\begin{figure}[t]
\centerline{
\epsfxsize = 0.5\textwidth \epsffile{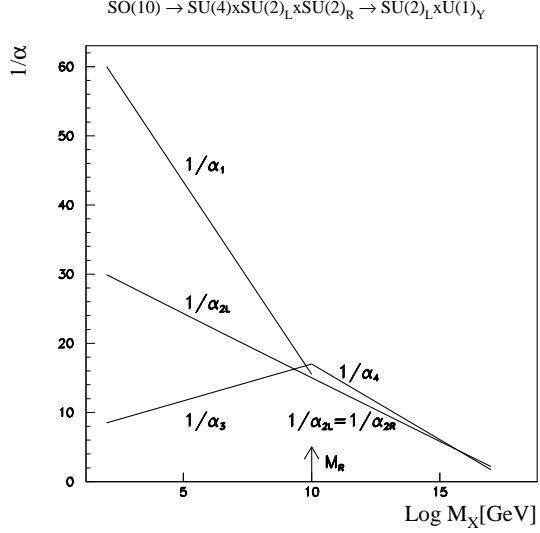} 
}
\caption{\it Running of the coupling constants assuming the 
intermediate Pati-Salam partial unification.
}
\label{fig:running}
\end{figure}

We shall exemplify a possible symmetry breaking path by assuming $SO(10)$
as grand unification group broken down to the SM symmetry in the
Pati--Salam chain\footnote{The evolution of couplings following 
such a chain, without additional light SUSY particles however, 
has been studied in Ref.\cite{utpal}.}:
\bea
SO(10) &\longrightarrow\!\!\!\!\!\!\!\!\!\!\!^{M_U}&  
SU(4)\times SU(2)_R\times SU(2)_L \nn \\
&\longrightarrow\!\!\!\!\!\!\!\!\!\!^{M_R}&  
SU(3)_c\times  SU(2)_L\times U(1)_{Y} \nn \\
&\longrightarrow\!\!\!\!\!\!\!\!\!\!\!^{M_W}&  
SU(3)_c\times U(1)_{em} 
\eea 
At the scale $M_R$ the couplings must satisfy
the  boundary conditions for Pati-Salam partial unification:
\bea
\alpha^{-1}_{3}(M_R)&=&\alpha^{-1}_{4}(M_R) \nn \\
\alpha^{-1}_{2}(M_R)&\equiv&\alpha^{-1}_{2L}(M_R)=\alpha^{-1}_{2R}(M_R) \nn \\
\alpha^{-1}_{1}(M_R)&=&\frac{3}{5}\alpha^{-1}_{2}(M_R)+
\frac{2}{5}\alpha^{-1}_{3}(M_R)
\eea
While the first two conditions are self--evident, the third condition 
follows from the breaking mechanism $R\times(B-L)\to Y.$
The partial unification scale is fixed by the third condition.
The low energy particle spectrum at scales $M_X$ below $M_R$ 
is already specified in \eq{spot1}. 
The corresponding SUSY beta functions $b_i$ are given by
\bea
b_1 &=& 
2N_F+\frac{3}{10}N_D+\frac{9}{5}N_\Delta +\frac{12}{5} N_\delta \nn \\
b_{2L} &=& 
-6 + 2N_F+\frac{1}{2}N_D+2N_\Delta  \nn \\
b_3 &=& 
-6 + 2N_F
\eea
where  $N_F=3$ is the number of families, and  
$N_D,$ $N_\Delta$ and $N_\delta$
are the numbers of doublet, triplet and singlet 
Higgs superfields, respectively; in the present example,
 $N_D=N_\Delta=N_\delta=2.$
The additional light Higgs superfields increase the slopes of $\alpha_1^{-1}$
and $\alpha_2^{-1},$ and they accelerate the running of the two couplings.
Above $M_R,$ the minimal $SU(2)_{L,R}$ superfield content
is naturally assumed to consist of two doublets, and of 
two left--handed and two right--handed triplets [$\Delta/\bar\Delta$ and
$\delta/\bar\delta,$ respectively]. 
This spectrum implies for the beta functions $b_{2L,R}$
\bea
b_{2L}=b_{2R}=
-6 + 2N_F+\frac{1}{2}N_D+2N_{\Delta,\delta}  
\eea
$\alpha_{2L}^{-1}=\alpha_2^{-1}$ therefore evolves with $M_X$
without any break across the scale $M_R$, cf. Fig.\ref{fig:running}.
Since the coupling 
$\alpha_{3}^{-1}$ becomes larger than $\alpha_{2}^{-1}$ near
$M_R$, the asymptotically free color $SU(3)_c$ sector must
transmute into an asymptotically non--free Pati--Salam $SU(4)$ sector
at $M_R$ in order to evolve into a grand unification crossing point 
with $SU(2)_L\times SU(2)_R$ at a large scale $M_U\gsim 10^{16}$ GeV.
This can be achieved by introducing 
$N_{10}$ ten-plet superfields, giving rise to the beta function
\bea
b_4=-12+2N_g+3N_{10}
\eea

Four such ten--plets are needed at least, in the present example,
to reach a unification point below the Planck scale;
with $N_{10}=4$ the $SO(10)$ unification point $M_U$ is located at a scale
$M_U\sim 10^{16}$ GeV while the LR symmetry breaking scale is 
$M_R\sim 10^{10}$ GeV, given the standard low energy couplings at $M_Z.$

Even though this specific example may look somewhat
baroque, as a result of the little motivated ten--plet spectrum,
it demonstrates nevertheless that light Higgs superfields can be accommodated
in grand--unification $SO(10)$ scenarios indeed. \\

%%%%%%%%%%%%%%%%%%%%%%%%

The two--component mass terms for the doubly charged higgsinos are 
derived from the superpotential \rfn{spot},
\bea
{\cal L}_{mass}=-M_{\tilde{\Delta}}\tilde{\Delta}^{++}_L
\tilde{\bar\Delta}^{--}_L- 
M_{\tilde{\delta}} \tilde{\delta}^{--}_R\tilde{\bar\delta}^{++}_R + h.c.
\eea
where the tilde denotes the fermionic component of the corresponding
superfield in \eq{fields}.
It follows from the above Lagrangean 
that the fermionic components of the two triplet 
(and singlet) superfields  combine to form one four--component 
fermion Dirac field; $\tilde{\Delta}^{++}_L$ and 
$(\tilde{\bar\Delta}^{--}_L)^c=\tilde{\bar\Delta}^{++}_R$ 
can be identified as left-- and right--chiral components
of one Dirac field $\tilde{\Delta}^{++}$ [$\tilde\delta$ correspondingly].
Therefore the left- and right-chirality components of the four--component
fermions, $\Dep_{L,R}$ and $\dep_{L,R},$ 
carry the same $SU(2)_L$ iso--quantum numbers
and they  couple not only to the photon but also to 
the $Z$-boson in exactly  the same way. 

The gauge interactions of the new higgsinos are described by 
the usual Lagrangean
\bea
{\cal L}= \overline{\Dep_L}i\!\!\not\!\!D\Dep_L +  
\overline{\Dep_R}i\!\!\not\!\!D\Dep_R +  
\overline{\dep_L}i\!\!\not\!\!D\dep_L +  
\overline{\dep_R}i\!\!\not\!\!D\dep_R
\label{charges} 
\eea 
where the covariant derivative is given by
$iD_\mu=i\partial_\mu+eQ_\gamma A_\mu +g_Z Q_ZZ_\mu $
with $g_Z=e/(s_W c_W).$ $Q_\gamma$ is the electric charge 
related to isospin $I_3$ and hypercharge $Y$ 
by the Gell--Mann--Nishijama relation $Q_\gamma=I_3+Y/2$
while the $Z$--charge follows from
$  Q_Z=I_3-s^2_W Q_\gamma $ 
with $s^2_W=1-c^2_W=\sin^2\theta_W$ being the weak 
mixing angle.
 Both left-- and right--chiral components of 
$\Dep$ and $\tilde\delta^{++}$ carry the same  isospin  $I_3=+1$ and
$I_3 = 0$, respectively.
The electroweak gauge theory of these fields is of vector--like
character.\\

The relevant Yukawa interactions for the doubly charged higgsinos
are given by the four--component Lagrangean
\bea
{\cal L}_Y=-2 f_\Delta \bar l^c_L\Dep_L \sle_L -2 f_\delta
\bar l^c_R\dep_R \sle_R + h.c. 
\label{yuk}
\eea
where the  subscripts $L,R$ denote the chirality of the fermions and
the type of the sleptons at the same time.

It is instructive to analyze also  the chargino and neutralino sectors
of the model {\it in toto}. Due to the two new triplets,
three charginos are generated, 
\bea
\psi^+&=&\left(-i\tilde\omega^+, \tilde h_2^+, \tilde\Delta^+ \right) \\
\psi^-&=&\left(-i\tilde\omega^-, \tilde h_1^-, \tilde{\bar\Delta}^-
\right)
\eea
and six two component neutralinos,
\bea
\psi^0=\left(-i\tilde b^0,-i\tilde\omega_3^0, \tilde h_1^0,\tilde h_2^0,
 \tilde\Delta^0,\tilde{\bar\Delta}^0 \right)
\eea
However, the new states do not mix with the MSSM states. Because there is 
no doublet-triplet mixing in the superpotential, the doublet 
higgsinos do not mix with the triplet higgsinos. Triplets can,
in principle, mix with gauginos due to the gauge-matter interactions 
\bea
{\cal L}_{int}=ig_a\sqrt{2}T^a_{ij}\phi_i^*\tilde\omega^a\psi_j + h.c.
\eea
where $T^a$ are the gauge group generators. However, the gaugino-triplet
higgsino mixing  terms are proportional to the vacuum expectation value
$v_L$  of the neutral left-handed triplet Higgs field $\Delta^0$ 
which is strongly constrained from the measurement of the $\rho$ parameter 
to be below 1 GeV. Therefore these mixing terms are negligible and the 
triplet higgsino decouples from the MSSM states.
The only mass term for the triplet neutralino generated by the 
superpotential \rfn{spot} 
is of the type $\Delta^0\bar\Delta^0$ which implies that
the components form one neutral Dirac fermion while the MSSM neutralinos
are
in general Majorana states.
As a result, the interactions of the triplet higgsinos are limited
to the Yukawa interactions of the type  \eq{yuk} and to the gauge 
interactions. They do not mix with the MSSM states and
the properties of the genuine charginos and neutralinos in
 the MSSM  are not modified.

%%%%%%%%%%%%%%%%%%%%%%%%%%%%%%%%%%%%%%%%%%%%%%%%%%
\subsection*{3. Production of  Doubly Charged Higgsinos}
\label{sec:3}
%%%%%%%%%%%%%%%%%%%%%%%%%%%%%%%%%%%%%%%%%%%%%%%%%%

The matrix elements of the processes \rfn{DD} and \rfn{dd}
can, quite generally, be expressed in terms of four 
bilinear charges \cite{R7,R7A}, classified according to the chiralities
$\alpha,
\beta=L,R$ of the associated lepton and higgsino currents, 
\begin{eqnarray}
T\left(\ee\ra\Dep\Dem\right)
 = \frac{e^2}{s}Q_{\alpha\beta}
   \left[\bar{v}(e^+)  \gamma_\mu P_\alpha  u(e^-)\right]
   \left[\bar{u}(\Dep) \gamma^\mu P_\beta 
               v(\Dem)\right]
\label{ampl}
\end{eqnarray}
and analogously for the process $\ee\ra\dep\dem .$
In this notation $\Dep,$ $\dep$ are defined as
particles and $\Dem,$ $\dem$  as antiparticles.
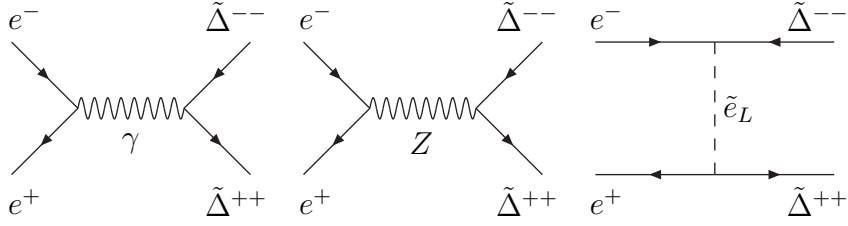
\begin{figure}[t]
\begin{center}
\begin{picture}(330,100)(0,0)
%1st diagram 
\Text(15,85)[]{$e^-$}
\ArrowLine(10,75)(35,50)
\ArrowLine(35,50)(10,25)
\Text(15,15)[]{$e^+$}
\Photon(35,50)(75,50){4}{8}
\Text(55,37)[]{$\gamma$}
\ArrowLine(100,75)(75,50)
%\Photon(75,50)(100,75){3}{7}
\Text(95,85)[]{$\Dem$}
\ArrowLine(75,50)(100,25)
%\Photon(100,25)(75,50){3}{7}
\Text(95,15)[]{$\Dep$}
% 2nd Diagram
\Text(125,85)[]{$e^-$}
\ArrowLine(120,75)(145,50)
\Text(125,15)[]{$e^+$}
\ArrowLine(145,50)(120,25)
\Photon(145,50)(185,50){4}{8}
\Text(165,37)[]{$Z$}
\ArrowLine(210,75)(185,50)
%\Photon(185,50)(210,75){3}{7}
\Text(205,85)[]{$\Dem$}
\ArrowLine(185,50)(210,25)
%\Photon(210,25)(185,50){3}{7}
\Text(205,15)[]{$\Dep$}
% 3rd Diagram
\Text(235,85)[]{$e^-$}
\ArrowLine(230,75)(275,75)
\Text(235,15)[]{$e^+$}
\ArrowLine(275,25)(230,25)
%\DashArrowLine(275,25)(275,75){5}
\DashLine(275,25)(275,75){5}
\Text(285,50)[]{$\se_L$}
\ArrowLine(320,75)(275,75)
%\Photon(275,75)(320,75){3}{7}
\Text(315,85)[]{$\Dem$}
\ArrowLine(275,25)(320,25)
%\Photon(320,25)(275,25){3}{7}
\Text(315,15)[]{$\Dep$}
\end{picture}
\end{center}
\caption{\it Feynmam graphs for the production of $\Dep\Dem$ 
 pair. The same set of diagrams describes $\dep\dem$ pair production with
$\se_L\to\se_R$. }
\label{fig:production}
\end{figure}

The process $\ee\ra\Dep\Dem$ is built up by 
$s$--channel $\gamma$ and $Z$ exchanges, and $t$--channel 
$\se_L$ exchange.  The corresponding  Feynman diagrams 
are depicted in Fig.\ref{fig:production}. 
After the appropriate  Fierz transformation,
 also the  $t$--channel amplitude can be cast in 
the form of \eq{ampl}. The charges for the process \rfn{DD} are 
given by
\begin{eqnarray}
\ee \ra \Dep\Dem: \,\,\;\;\;\;
Q_{LL}&=&1+ 2 \cot^2 2\theta_W D_Z
          + 2(f_\Delta^2/e^2) D_{\se_L}
         \nonumber\\ 
Q_{LR}&=&1+ 2 \cot^2 2\theta_W D_Z 
         \nonumber\\
Q_{RR}&=& Q_{RL}\;=\; 1- (\cos 2\theta_W/\cos^2\theta_W) D_Z
\label{qD}
\end{eqnarray}
The first index in $Q_{\alpha \beta}$ refers to the chirality of the
$e^\pm$ 
current, the second index to the chirality of the
$\tilde{\Delta}^{\pm\pm}$ 
current. In the process \rfn{DD} the $\tilde{e}_L$ 
exchange affects only the $LL$ chirality charge while all other amplitudes 
are built up by $\gamma$ and $Z$ exchanges. $D_{\tilde{e}}$ denotes the
slepton propagator $D_{\tilde{e}} = s/(t- m_{\tilde{e}}^2)$, and 
$D_Z$ the $Z$ propagator $D_Z=s/(s-m^2_Z+im_Z\Gamma_Z)$; the non--zero 
$Z$ width 
can in general be neglected for the energies considered in the present 
analysis so that the charges are real. 
\begin{figure}[t]
\centerline{
\epsfxsize = 0.5\textwidth \epsffile{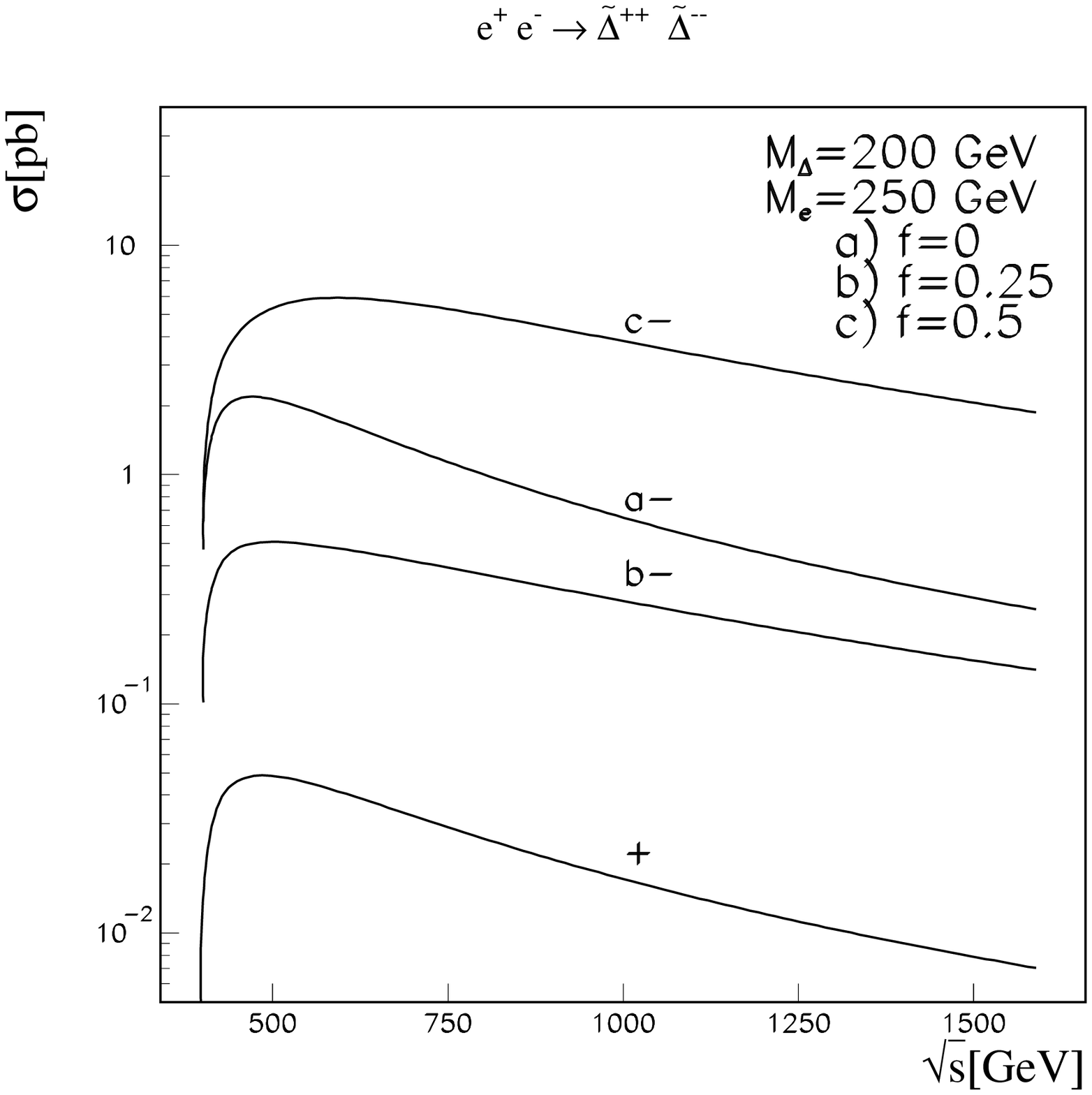} 
\hfill
\epsfxsize = 0.5\textwidth \epsffile{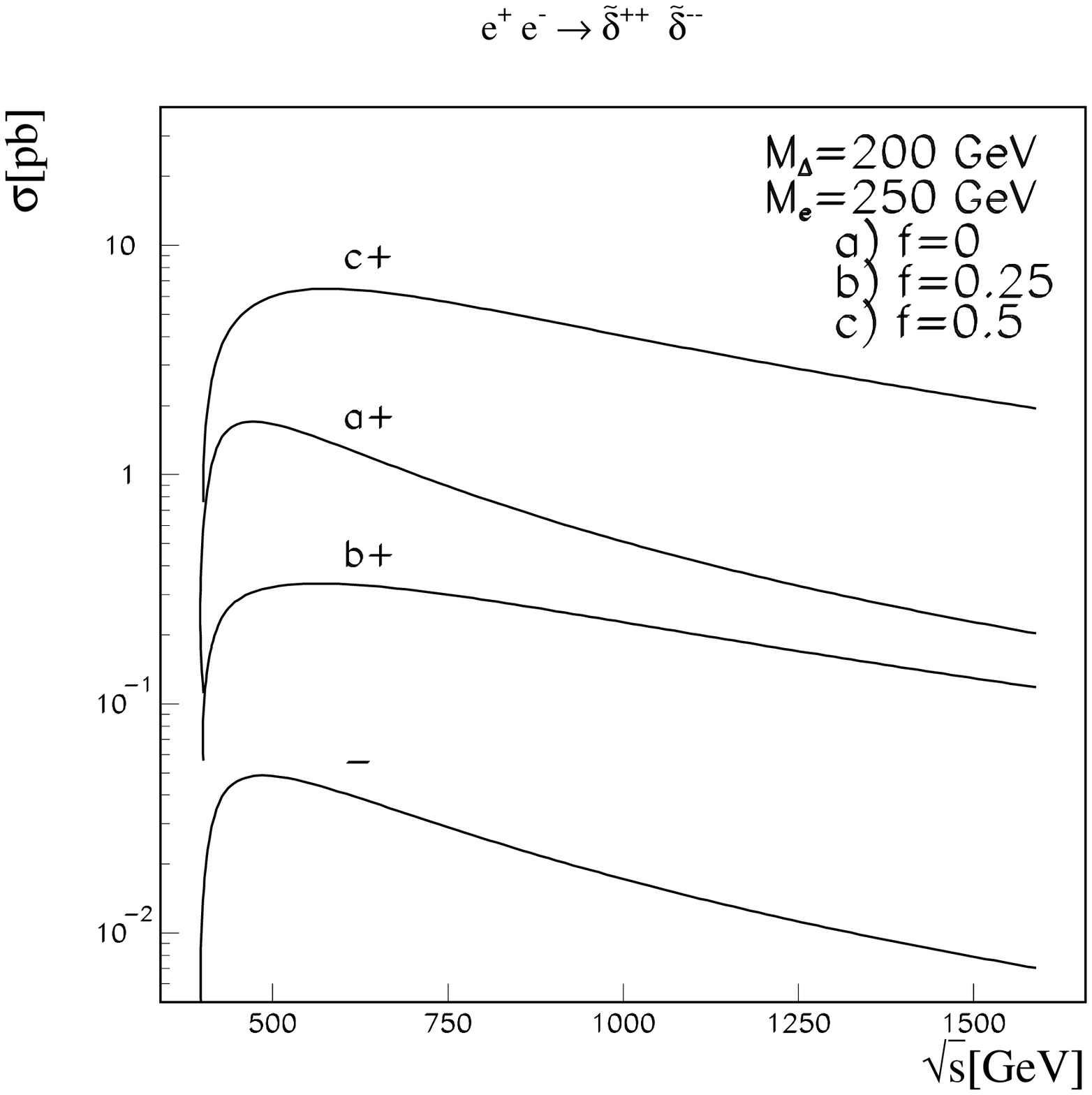}
}
\caption{\it The polarized cross sections for the pair production of 
            doubly charged higgsinos $\tilde\Delta^{\pm\pm}$ and 
$\tilde\delta^{\pm\pm}$ 
as functions of the  collision energy. The
            choice of the masses, the Yukawa couplings and the electron 
beam polarizations (L/R=-/+) are indicated in the figures.}
\label{fig:cre}
\end{figure}

In contrast to \rfn{DD}, the process \rfn{dd}
describes the production of a  right-handed  singlet. 
The $\dep$  coupling to the $Z$ boson is modified according to \eq{charges}
and the $t$--channel exchange graph in Fig.\ref{fig:production} 
involves the right--chiral selectron $\se_R.$ 
The corresponding charges for the process  $\ee\ra\dep\dem$ 
are given by
\begin{eqnarray}
\ee \ra \dep \dem: \;\;\;\;\;
Q_{LL}&=&Q_{LR}\;=\; 1-(\cos 2\theta_W/\cos^2\theta_W) D_Z  \nonumber\\
 Q_{RL} &=& 1+ 2\tan^2\theta_W D_Z   \nonumber\\
 Q_{RR}&=&1 +2\tan^2\theta_W D_Z+ 2(f_\delta^2/e^2) D_{\se_R}
\label{qd}
\end{eqnarray}
As predicted by the chirality of $\dep$, 
the non-photonic contribution from the $t$--channel $\se_R$ exchange
affects
only the $RR$ chirality charge. 

To derive a transparent form for cross sections and
polarization vectors, the following  quartic charges are generally 
introduced,
\begin{eqnarray}
&&    Q_1=\frac{1}{4}\left[|Q_{RR}|^2+|Q_{LL}|^2
                          +|Q_{RL}|^2+|Q_{LR}|^2\right] \nonumber\\
&&    Q_2=\frac{1}{2}{\rm Re}\left[Q_{RR}Q^*_{RL}
                                  +Q_{LL}Q^*_{LR}\right] \nonumber\\
&&    Q_3=\frac{1}{4}\left[|Q_{RR}|^2+|Q_{LL}|^2
                          -|Q_{RL}|^2-|Q_{LR}|^2\right] 
\end{eqnarray}
and
\begin{eqnarray}
&&   Q'_1=\frac{1}{4}\left[|Q_{RR}|^2+|Q_{RL}|^2
                          -|Q_{LR}|^2-|Q_{LL}|^2\right] \nonumber\\
&&   Q'_2=\frac{1}{2}{\rm Re}\left[Q_{RR}Q^*_{RL}
                                  -Q_{LL}Q^*_{LR}\right] \nonumber\\
&&   Q'_3=\frac{1}{4}\left[|Q_{RR}|^2+|Q_{LR}|^2
                          -|Q_{RL}|^2-|Q_{LL}|^2\right]
\end{eqnarray}
which describe the independent experimental observables.

The final state probability may be expanded in terms of the unpolarized
cross 
section, the polarization vectors of $\Dep$ and $\Dem$, 
and the spin--spin correlation tensor. 
The $\Dep$ production angle, with respect to the electron
flight--direction,
will be denoted by $\Theta$.
Defining the $\hat{z}$ axis by the $e^-$ 
momentum, the $\hat{x}$ axis in the production plane with $\hat{x} 
\cdot \vec{p}_{\Dep} >0,$ and $\hat{y} =\hat{z}\times\hat{x}$ 
in the rest frame of the charginos, cross section and spin--density
matrices are defined  as \cite{R9}: 
\begin{eqnarray}
&&\frac{{\rm d}\sigma}{{\rm d}\cos\Theta}
     (\lambda\lambda';\bar{\lambda}\bar{\lambda}')
 = \\
&& { }\hskip 0.2cm =\frac{{\rm d}\sigma}{{\rm d}\cos\Theta}\frac{1}{4}
   \bigg[(I)_{\lambda'\lambda}(I)_{\bar{\lambda}\bar{\lambda}'}
  +{\cal P}_i (\tau^i)_{\lambda'\lambda}(I)_{\bar{\lambda}\bar{\lambda}'}
  +\bar{\cal P}_i
(I)_{\lambda'\lambda}(\tau^i)_{\bar{\lambda}\bar{\lambda}'}
  +{\cal Q}_{ij}
(\tau^i)_{\lambda'\lambda}(\tau^j)_{\bar{\lambda}\bar{\lambda}'}
   \bigg]  \nonumber 
\end{eqnarray}
where $\lambda(\lambda')$ and $\bar{\lambda}(\bar{\lambda}') = \pm 1$ 
are twice the helicities of  $\Dep$ and $\Dem,$ and  
${\cal P}_i,$ $\bar {\cal P}_i$ are the components of the polarization  
vectors of $\Dep$ and $\Dem,$ respectively,
  with respect to the reference frame introduced 
above.  The tensor
${\cal Q}_{ij}$ denotes the spin--spin correlation matrix of the 
  $\Dep$ and $\Dem$ spins. The same decomposition can be 
formulated for the $\dep$ and $\dem$ pair.

\begin{figure}[t]
%\begin{center}
\centerline{
\epsfxsize = 0.5\textwidth \epsffile{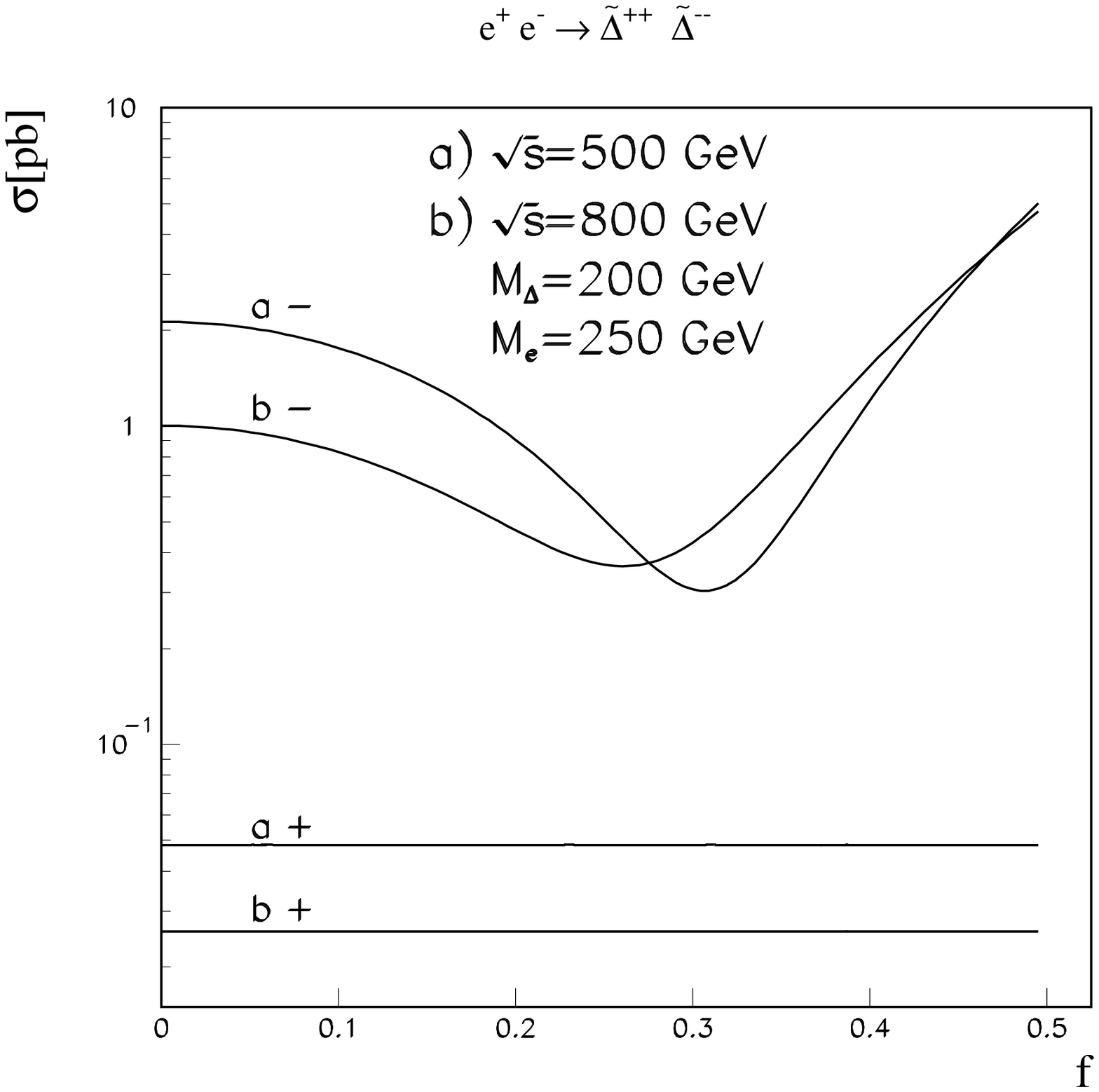} 
\hfill
\epsfxsize = 0.5\textwidth \epsffile{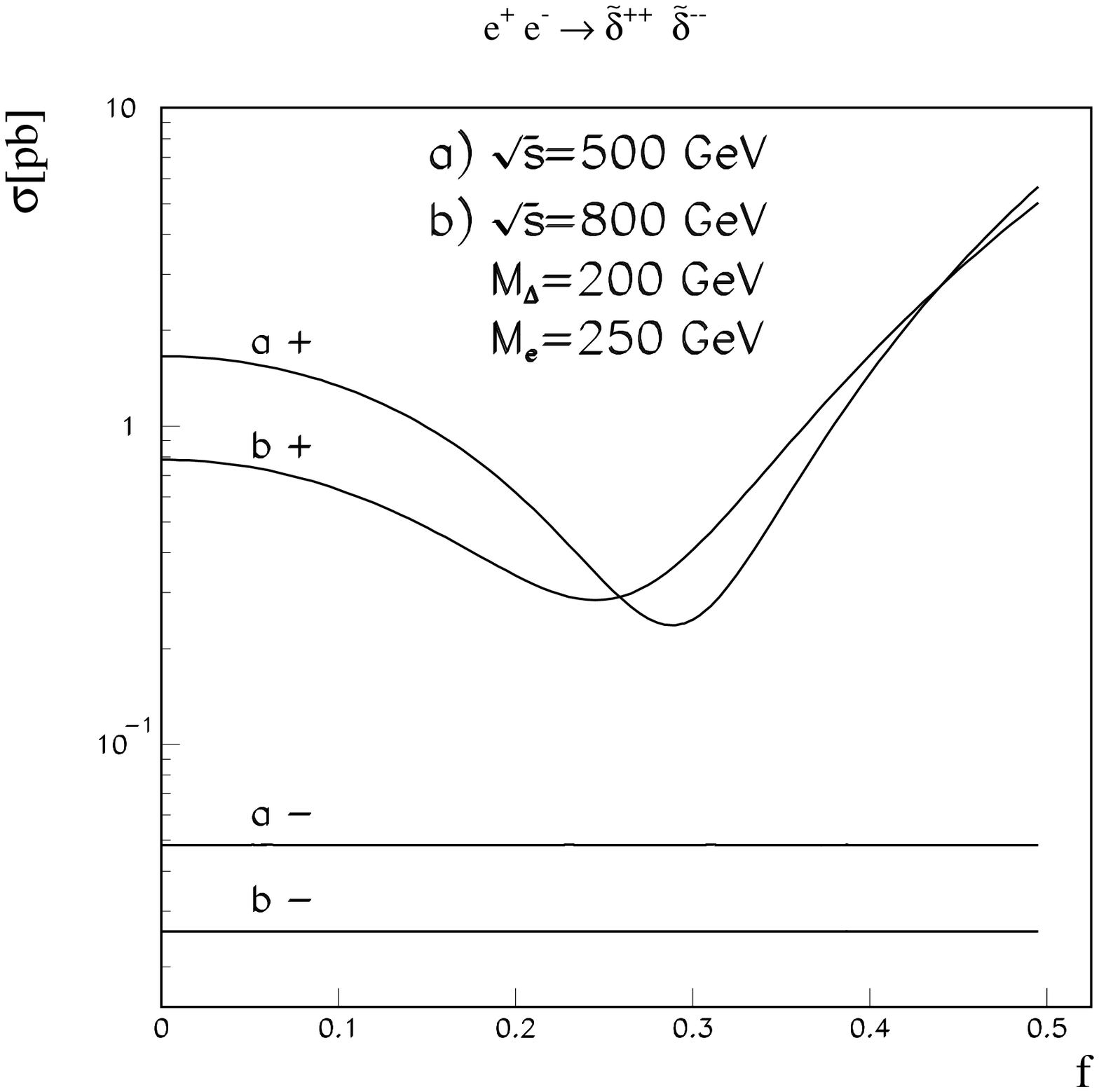}
}
%\hbox to\textwidth{\hss\epsfig{file=crf.ps,height=7cm}\hss}
%\vskip -1.cm
%\end{center}
\caption{\it The cross sections for the processes  
            $\ee\to\Dep\Dem$ and $\ee\to\dep\dem$
for polarized beams as  functions of the couplings  $f_\Delta$
and $f_\delta,$ respectively.
}
\label{fig:crf}
\end{figure}
%

%%%%%%%%%%%%%%%%%%%%%%%%%%%%%%%%%%%%%%%%%%%%%%%
\subsubsection*{3.1 The production cross section}
%%%%%%%%%%%%%%%%%%%%%%%%%%%%%%%%%%%%%%%%%%%%%%%

The cross sections of the processes \rfn{DD} and \rfn{dd} depend
on the parameters of the underlying theory: on the
masses of $\Dep,$ $\dep$,  on 
their couplings to the $Z$ boson,
and also  the strength and chirality of the  
$\Dep$ and $\dep$ couplings to electron--selectron pairs.

The unpolarized differential cross sections of the 
processes \rfn{DD}, \rfn{dd} are given by 
\begin{eqnarray}
\frac{{\rm d}\sigma}{{\rm d}\cos\Theta}
 =\frac{\pi\alpha^2}{2 s} \beta 
  \left\{(1+\beta^2\cos^2\Theta)Q_1+(1-\beta^2)Q_2
         +2\beta\cos\Theta Q_3\right\}
\label{crup}
\end{eqnarray}
If the beams are polarized, the same universal form holds for the
 cross section. The quartic charges must be adjusted however by 
restricting the sum to either $Q_{R\ast}$ or $Q_{L\ast}$ terms for 
right-- and left--handed electrons, respectively; moreover,
a factor 1/4 accounting for the spin average must be replaced, 
{\it e.g.,} by unity if the $e^{\pm}$ beams are both polarized.

Since the cross sections  are proportional to $\beta,$ it is possible
to
carry out a very precise determination of the 
 $\Dep,$ $\dep$ masses  at the 
production threshold \cite{R4} to an accuracy of $\sim$100 MeV. 
The threshold cross sections for 
the longitudinal 
electron beam polarizations are given by
\begin{eqnarray}
\sigma_R(\ee\ra\Dep\Dem)
& = &\frac{4\pi\alpha^2}{ s} \beta 
  \left[1-\left(\frac{\cos 2\theta_W}{\cos^2\theta_W} \right)
\frac{s}{s-M_Z^2}\right]^2 \nn \\
\sigma_L(\ee\ra\Dep\Dem)
& =&\frac{4\pi\alpha^2}{ s} \beta 
  \left[ 1+2\cot^2 2\theta_W
\frac{s}{s-M_Z^2}- 
\frac{f_\Delta^2}{e^2}\frac{4s}{s+4m^2_{\sle_L}}\right]^2 \nn \\
\sigma_R(\ee\ra\dep\dem)
& =&\frac{4\pi\alpha^2}{ s} \beta 
  \left[ 1+2\tan^2\theta_W  \frac{s}{s-M_Z^2}-
\frac{f_\delta^2}{e^2}\frac{4s}{s+4m^2_{\sle_R}}\right]^2 \nn \\ 
\sigma_L(\ee\ra\dep\dem)
& =&\frac{4\pi\alpha^2}{ s} \beta  \left[ 1- 
\left(\frac{\cos 2\theta_W}{\cos^2\theta_W} \right)
\frac{s}{s-M_Z^2}
 \right]^2
\end{eqnarray}
The angular distributions are isotropic at the thresholds.
The  cross sections   depend 
strongly on the beam polarization
and on the nature of the doubly charged particle.
\begin{figure}[t]
\centerline{
\epsfxsize = 0.5\textwidth \epsffile{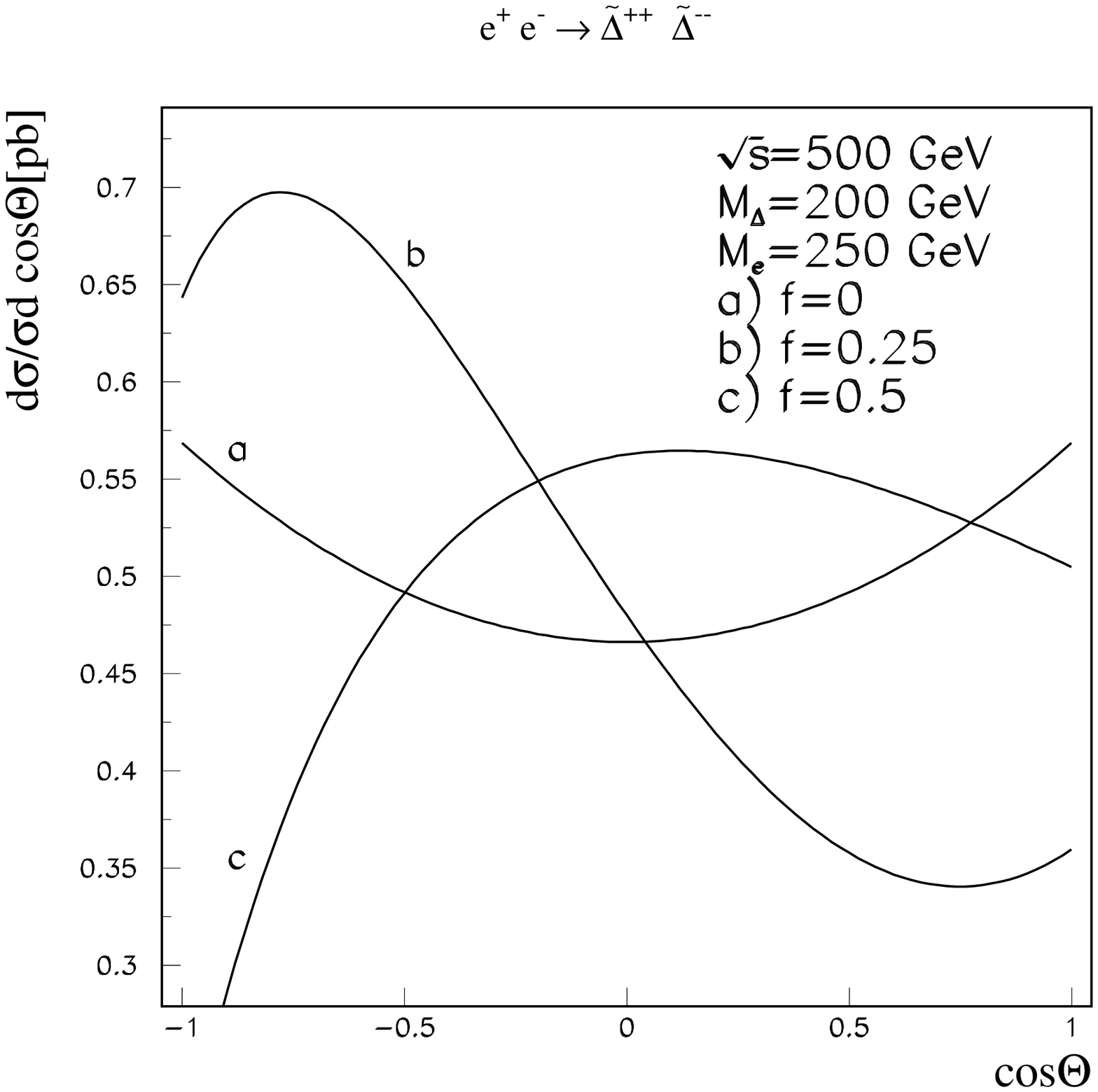} 
\hfill
\epsfxsize = 0.5\textwidth \epsffile{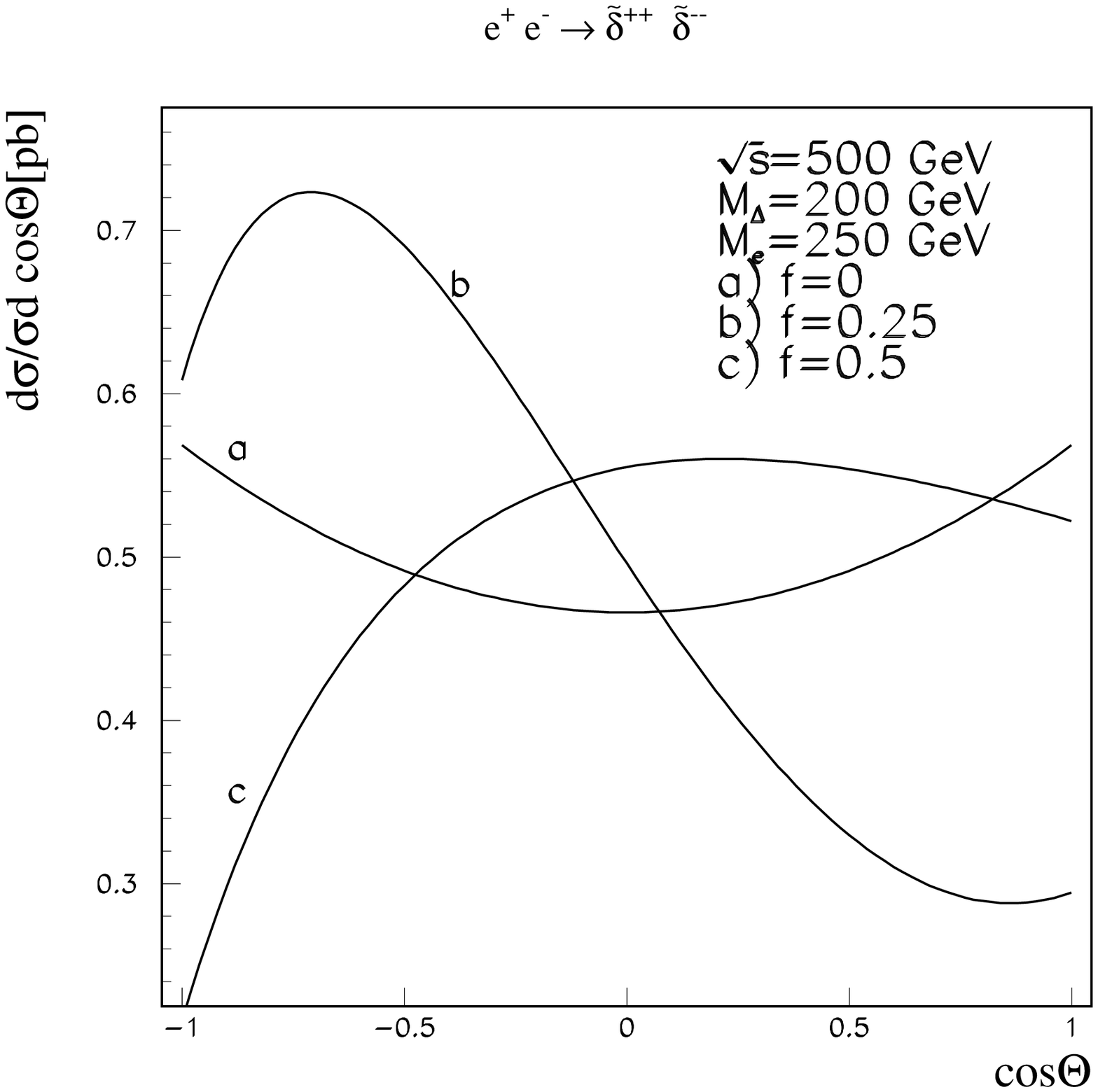}
}
\caption{\it The normalized differential cross sections (unpolarized beams)
 for             $\ee\to\Dep\Dem$ and $\ee\to\dep\dem$ 
   for different
            values of the couplings $f.$  }
\label{fig:dcrf}
\end{figure}

While the steep rise of the cross sections
near the thresholds can be exploited to determine
the masses very accurately, the charges and couplings of the 
particles can be measured at energies sufficiently above the 
thresholds. The relevant parameters are the 
isopin $I_{3}$ and the coupling $f$ between electron,
selectron and higgsino. The sensitivity to these parameters is
demonstrated in Figs.\ref{fig:cre} for a few examples.

Since $\Dem$ does not couple to right--handedly polarized electrons, 
$t$--channel selectron exchange does not contribute to
$\sigma_R (\tilde\Delta)$, and the cross section can be used
to measure the isospin $I_{3} (\Dep) = + 1$. The same 
holds for $\sigma_L (\tilde\delta)$ which 
determines $I_{3} (\dep) = 0$. The vector--character
of the $\tilde\Delta$ and $\tilde{\delta}-Z$ interactions
can be established experimentally by proving that the
forward--backward asymmetries of  $\Dep$ and $\dep$ vanish for
right-- and left--handedly polarized electron beams.

The mirror cross sections $\sigma_L (\tilde{\Delta})$ and 
$\sigma_R(\tilde{\delta})$ can subsequently
be used to determine the $e\tilde{e}\tilde{\Delta}$ and
$e\tilde{e}\tilde{\delta}$ couplings $f_{\Delta}$ and $f_{\delta}$.
The $f$ dependence of $\sigma_L (\tilde{\Delta})$ and 
$\sigma_R(\tilde\delta)$ is shown in Figs.\ref{fig:crf}. The
effect of the $t$--channel selectron exchange
is very important for couplings of the same
order as the electromagnetic coupling, $f \sim e \sim 1/3$. For a large
range of the parameter values, the couplings $f$ can be
derived from the measured cross section only up to a two--fold 
ambiguity. For large $f$ values, the solution is unique.

For asymptotic energies the contributions from
$s$-channel $\gamma, Z$ exchange and $t$--channel
$\tilde e_{L,R}$ exchange are of the same order:
\begin{eqnarray}
\sigma_R(\ee\ra\Dep\Dem)
& \Rightarrow&\frac{8\pi\alpha^2}{3 s}\tan^4\theta_W
 \nn \\
\sigma_L(\ee\ra\Dep\Dem)
& \Rightarrow & \frac{8\pi\alpha^2}{3 s} 
  \left[\frac{1}{4}\left( \tan^2\theta_W+\cot^2\theta_W \right)^2 \right. 
\nn \\
& &\left.
-\frac{3}{2}\frac{f_\Delta^2}{e^2}
\left( \tan^2\theta_W+\cot^2\theta_W \right)
+6\frac{f_\Delta^4}{e^4}\right]
\end{eqnarray}
and
\begin{eqnarray}
\sigma_R(\ee\ra\dep\dem)
& \Rightarrow &\frac{8\pi\alpha^2}{3 s} 
  \left[(1+2\tan^2\theta_W)^2-3\frac{f_\delta^2}{e^2}(1+2\tan^2\theta_W)
+6\frac{f_\delta^4}{e^4}\right]
 \nn \\
\sigma_L(\ee\ra\dep\dem)
& \Rightarrow &\frac{8\pi\alpha^2}{3 s} \tan^4\theta_W
\end{eqnarray}

Also
the angular distributions, Fig.\ref{fig:dcrf}, are sensitive
to the $f$ values. However, due to two neutralinos
escaping the detector, they cannot directly be used to
resolve the ambiguity; the detailed discussion of the 
resolution is deferred to Section 5.

%%%%%%%%%%%%%%%%%%%%%%%%%%%%%%%%%%%%%%%%%%%%%%%%%%%%%%%%%
\subsubsection*{3.2 The polarization of 
 doubly charged higgsinos}
\label{subsec:polarization vector}
%%%%%%%%%%%%%%%%%%%%%%%%%%%%%%%%%%%%%%%%%%%%%%%%%%%%%%%%%

The polarization vector 
$\vec{\cal P}=({\cal P}_T,{\cal P}_N,{\cal P}_L)$ 
is defined in the rest frame of the particles $\Dep$ and $\dep$. 
${\cal P}_L$ denotes the component parallel to the flight 
direction in the c.m. frame, ${\cal P}_T$ the component in the 
production plane, and ${\cal P}_N$ the  component normal to the production 
plane. 
\begin{figure}[t]
%\begin{center}
\centerline{
\epsfxsize = 0.5\textwidth \epsffile{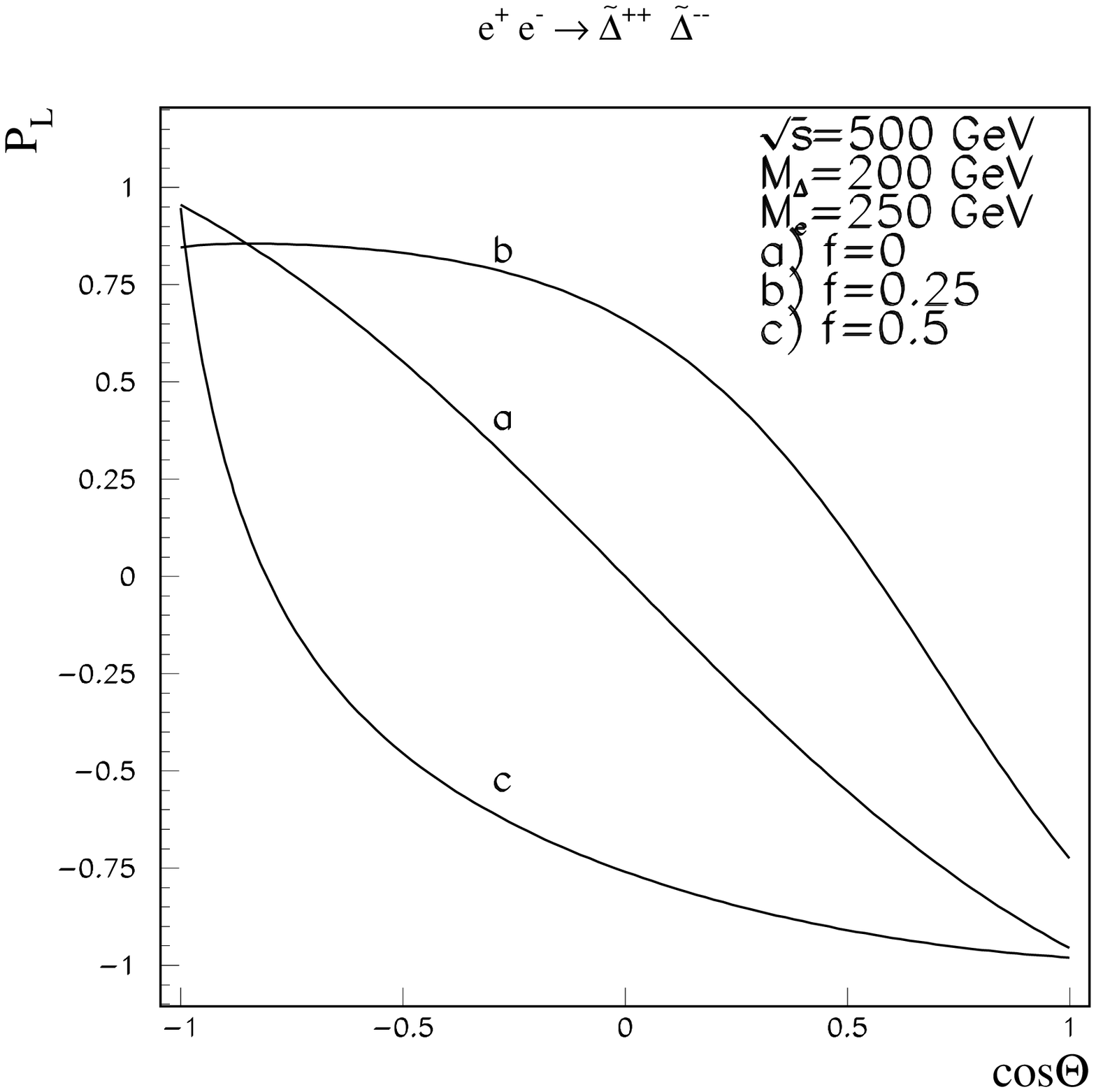} \hfill
\epsfxsize = 0.5\textwidth \epsffile{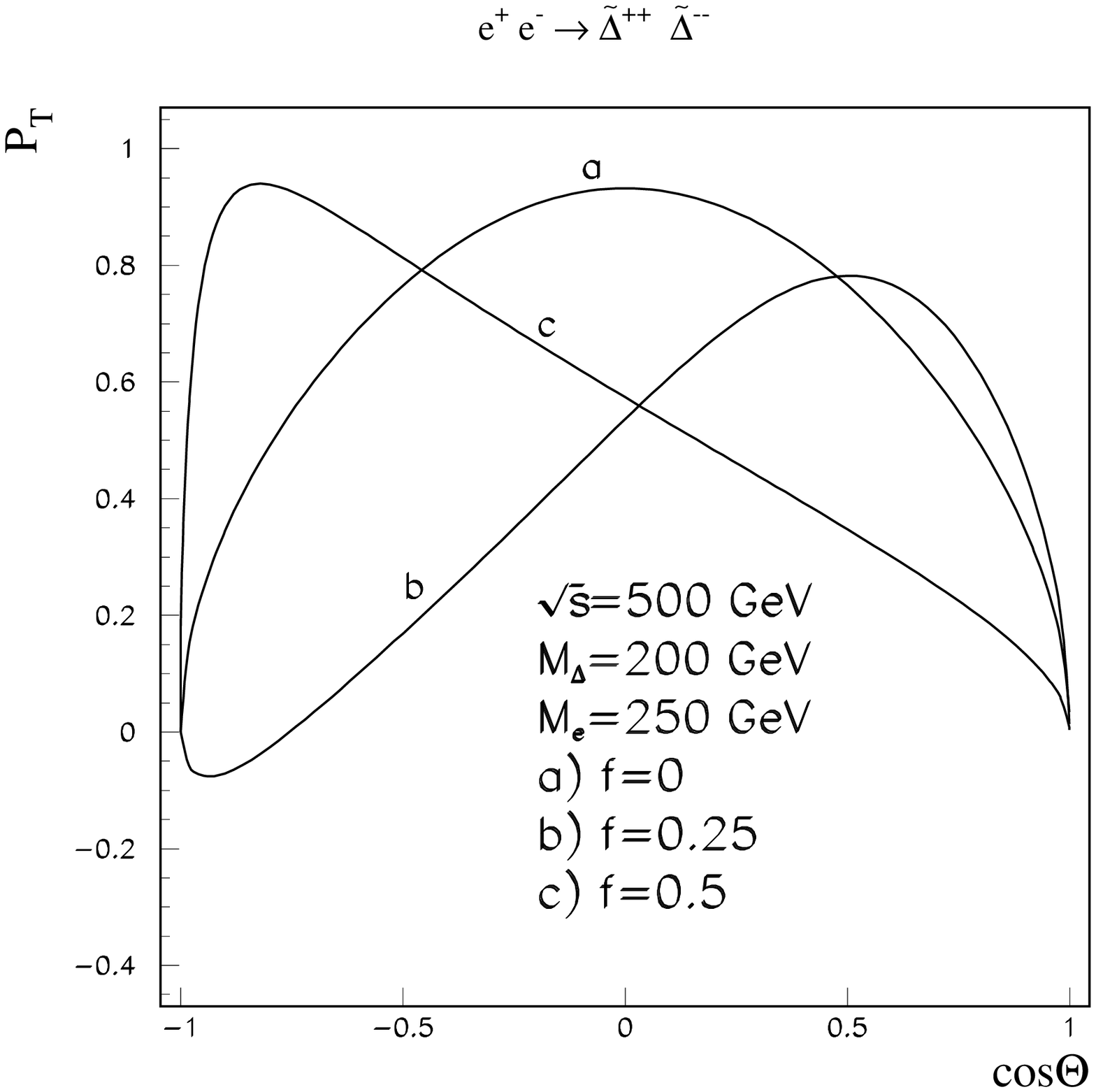}
}
%\hbox to\textwidth{\hss\epsfig{file=dpl.ps,height=7cm}\hss}
%\vskip -1.cm
%\end{center}
\caption{\it The  longitudinal and transverse
polarization components of 
           $\Dep$ produced with unpolarized beams for different
            values of the coupling $f.$  }
\label{fig:dpl}
\end{figure}

The normal component can only be generated by complex production
amplitudes.
Without loss of generality the couplings $f$ can be chosen  real.
The non--zero width of the $Z$ boson and loop corrections generate 
non--trivial phases; 
however, the width effect is negligible for high energies and the 
effects due to radiative corrections are small. Neglecting 
the small $Z$--width and the loop corrections,
 the normal polarizations of $\Dep,$
$\dep$  are vanishing. 
\begin{figure}[t]
%\begin{center}
\centerline{
\epsfxsize = 0.5\textwidth \epsffile{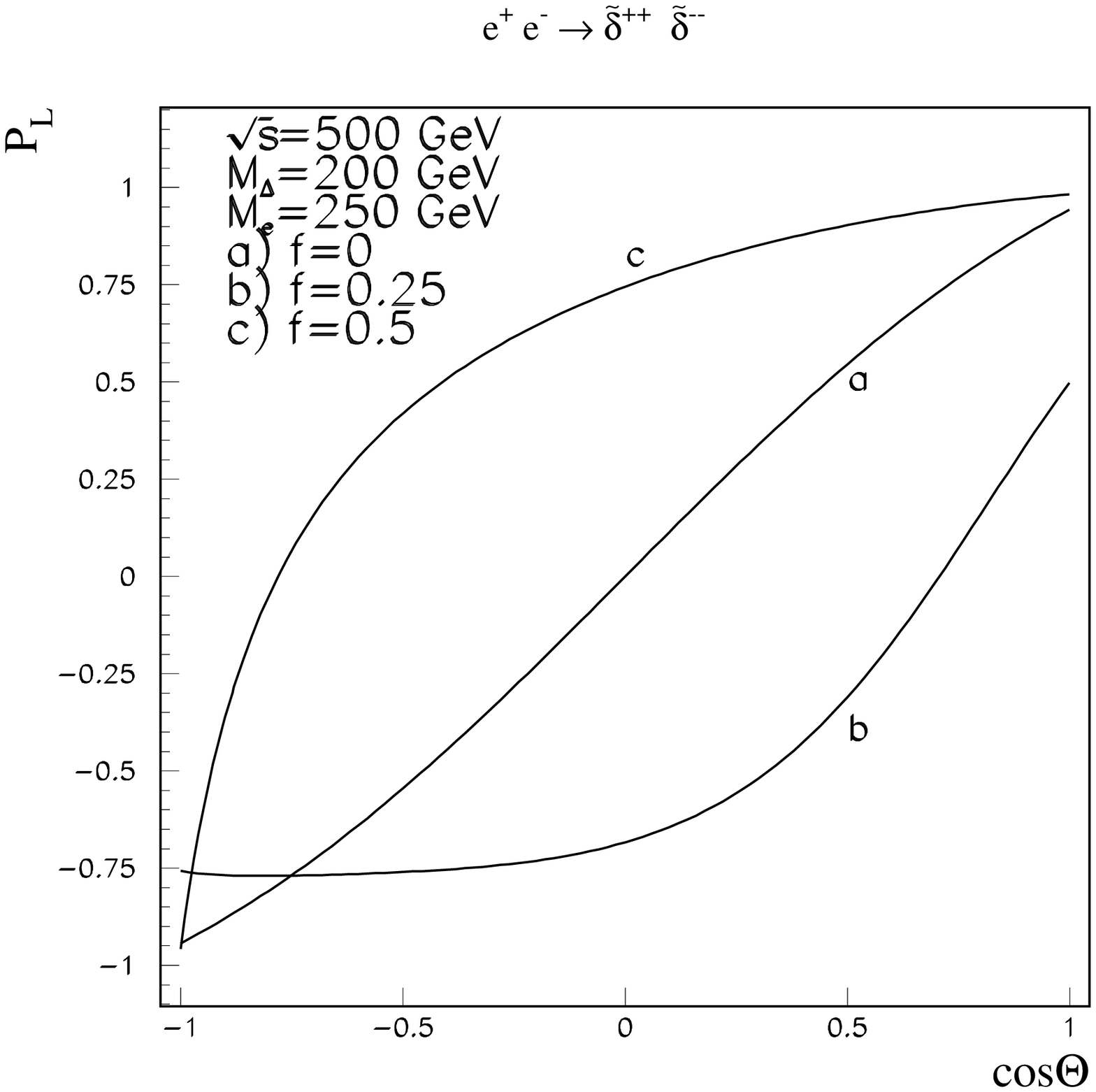} \hfill
\epsfxsize = 0.5\textwidth \epsffile{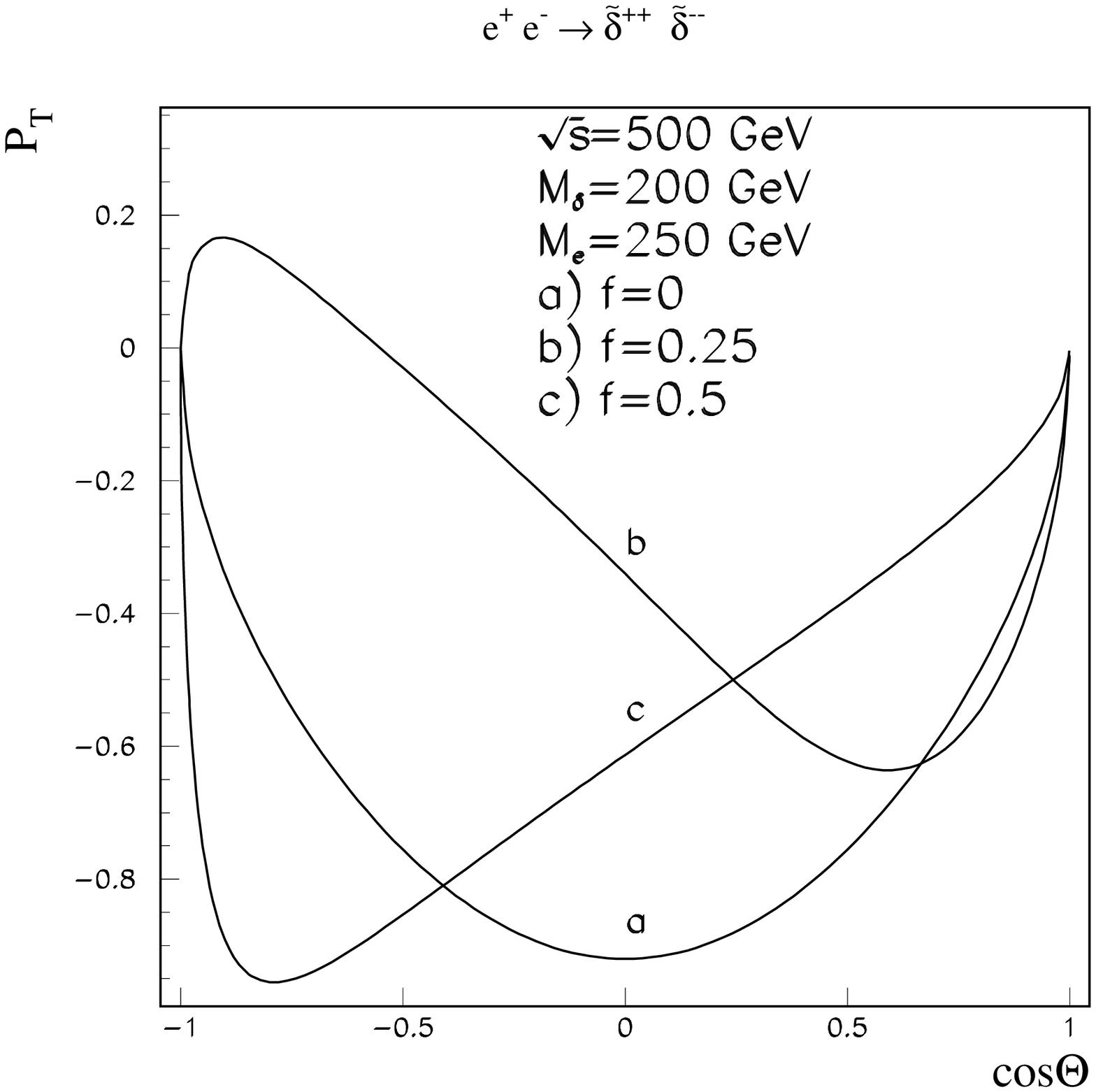}
}
%\hbox to\textwidth{\hss\epsfig{file=dpl.ps,height=7cm}\hss}
%\vskip -1.cm
%\end{center}
\caption{\it The  longitudinal and transverse
polarization components of 
           $\dep$ produced with unpolarized beams for different
            values of the couplings $f.$  }
\label{fig:dplr}
\end{figure}

In terms of the quartic charges the 
longitudinal and transverse components of the $\Dep$
polarization vector can be expressed as \cite{R7A}
\begin{eqnarray}
\label{PL}
&& {\cal P}_L    =4\left\{ (1+\beta^2)\cos\Theta{Q'}_1
                      +(1-\beta^2)\cos\Theta {Q'}_2
                      +(1+\cos^2\Theta)\beta {Q'}_3 \right\} / {\cal N}
 \\
&& {\cal P}_T  =-4\sqrt{1-\beta^2}\sin\Theta
                 \left\{ {Q'}_1+{Q'}_2+\beta\cos\Theta{Q'}_3
                 \right\}/{\cal N} 
\end{eqnarray}
where
\begin{eqnarray}
{\cal N}=4\{(1+\beta^2\cos^2\Theta)Q_1+(1-\beta^2)Q_2+2\beta\cos\Theta
Q_3\}
\end{eqnarray}
Close to the production threshold, 
${\cal P}_L$ and ${\cal P}_R$ are given by the same combination
of quartic charges:
\begin{eqnarray}
{\cal P}_L \rightarrow \frac{{Q'}_1 + {Q'}_2}{Q_1 + Q_2} \cos\Theta 
\makebox[6mm]{} \makebox{and} \makebox[6mm]{}
{\cal P}_T \rightarrow - \frac{{Q'}_1 + {Q'}_2}{Q_1 + Q_2} \sin\Theta
\end{eqnarray}

The values of the longitudinal polarization component ${\cal P}_L$
and the transverse component ${\cal P}_T$  of $\Dep$ and $\dep,$
are shown in Figs.\ref{fig:dpl} and \ref{fig:dplr}, 
respectively. While the curves denoted by 
$a$ ($f=0$)  in Fig.\ref{fig:dpl} are characteristic 
for $\gamma$ and $Z$ exchange in  production of chiral fermions, 
 the polarization curves  change  substantially for non--zero $f$.
The $\Dep$ and $\dep$ polarizations affect the decay
distributions so that polarization effects can serve as one of the tools
to resolve the two--fold ambiguity in the
measurements of the couplings $f$.

\subsubsection*{3.3 Production of doubly charged higgsinos in 
$\gamma\gamma$ collisions}

\begin{figure}[t]
\centerline{
\begin{picture}(150,100)(-25,-60)
\Photon(0,90)(75,90){4}{5}
\Photon(0,10)(75,10){4}{5}
\ArrowLine(75,90)(150,90)
\ArrowLine(150,10)(75,10)
\ArrowLine(75,10)(75,90)
\Text(30,80)[]{$\gamma$}
\Text(30,20)[]{$\gamma$}
\Text(120,80)[]{$\Dep$}
\Text(90,50)[]{$\Dep$}
\Text(120,22)[]{$\Dem$}
\end{picture}
\hfill
\epsfxsize = 0.5\textwidth \epsffile{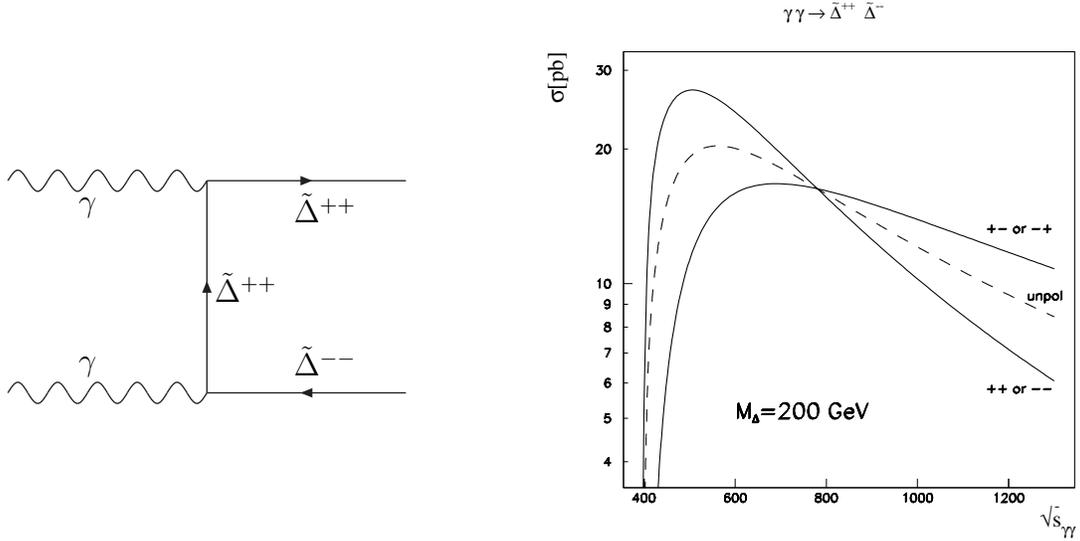}
}
\caption{\it 
 The production cross sections of the higgsinos $\Dep,\dep$ in the 
      $J_z=0$ and 2 channels, as well as the unpolarized cross section,
      in $\gamma\gamma$ collisions as a function of the 
invariant $\gamma\gamma$
collision  energy for fixed higgsino mass $M_{\Dep}=200$ GeV.
 }
\label{fig:gg}
\end{figure}

The double electric charges render $\gamma\gamma$ collisions an interesting
channel for the production of the higgsinos $\Dep,\dep$:
\bea
             \gamma\gamma\to\Dep\Dem\;\; \mrm{and}\;\;\,\dep\dem
\eea                             
The cross section increases by a factor $2^4=16$ compared to the 
$\gamma\gamma$ 
production of singly charged fermions of the same mass. Moreover,
for a given mass the theoretical prediction of the cross section is
parameter-free. Quasi-monoenergetic $\gamma\gamma$ collisions can be generated
by Compton back-scattering of laser light \cite{telnov}
if the conversion points and the collision point are slightly separated.
The $\gamma\gamma$ cm energy amounts to a fraction  0.8 of the initial 
$e^+e^-$ cm energy. 
 
For $J_z=0$ and $J_z=2$ $\gamma\gamma$ states, the production 
cross sections can be adapted from Ref.\cite{gg} by taking into account the
double electric charge of higgsinos:
\bea
\sigma(\gamma_\pm\gamma_\pm\to\Dep\Dem)
&=&\frac{32\pi\alpha^2}{s}
\left[ 2\beta\left(1+\beta^2\right)+
 \left(1-\beta^4\right)
\ln\frac{1+\beta}{1-\beta} \right] \nn \\
\sigma(\gamma_\pm\gamma_\mp\to\Dep\Dem)
&=&\frac{32\pi\alpha^2}{s}
\left[ -2\beta\left(5-\beta^2\right)+ 
\left(5-\beta^4\right)
\ln\frac{1+\beta}{1-\beta} \right] 
\eea
adding up to the unpolarized cross section 
\bea
\sigma(\gamma\gamma\to\Dep\Dem)
&=&\frac{32\pi\alpha^2}{s}
\left[ 2\beta\left(-2+\beta^2\right)+
 \left(3-\beta^4\right)
\ln\frac{1+\beta}{1-\beta} \right] 
\eea
 $s$ is the photon--photon collision energy squared and 
$\beta=\sqrt{1-4M_{\Dep}^2/s}$ the velocity of the 
higgsinos.
The angular distribution of the  higgsinos
in the $\gamma\gamma$ c.m. frame is given by
\bea
\frac{d\sigma}{d\cos\Theta}(\gamma\gamma\to\Dep\Dem) 
= \frac{16\pi\alpha^2}{s}
\left[-2+s(s+4M_{\Dep}^2)D-
4M_{\Dep}^4s^2D^2  \right]
\eea 
 with $D^{-1}=(t-M^2_{\Dep})(u-M^2_{\Dep})$ 
and
$t(u)-M^2_{\Dep}=s(1\mp\beta\cos\Theta)/2$
being the momentum transfer.
The form of the cross section is the same for $\dep$ production.

Typical examples (see also Ref.\cite{higgsino2}) 
are displayed in Fig.\ref{fig:gg}. For
asymptotic energies the $J_z=2$ channel is dominant, as  well-known. For
moderate energies however the role is reversed and $J_z=0$ is the leading
channel. Near the threshold, the cross sections rise steeply proportional
to the velocity $\beta$. 
As expected, with $\sigma_{max}\approx25$ pb the cross
sections are very large indeed, resulting in $2.5\times10^6$ events at
an integrated luminosity of 100 $fb^{-1}$, \ie\ about one fifth of the 
$e^+e^-$ collider luminosity. 
This event rate allows to perform detailed analyses
of the higgsino $\Dep,\dep$ decays, which this way can solidly be based 
on a parameter-free production mechanism.

%%%%%%%%%%%%%%%%%%%%%%%%%%%%%%%%%%%%%%%%%%%%%%%%%%
\subsection*{4. Doubly Charged Higgsino Decays}
\label{sec:decay}
%%%%%%%%%%%%%%%%%%%%%%%%%%%%%%%%%%%%%%%%%%%%%%%%%%

\begin{figure}[t]
%\begin{center}
\centerline{
\epsfxsize = 0.5\textwidth \epsffile{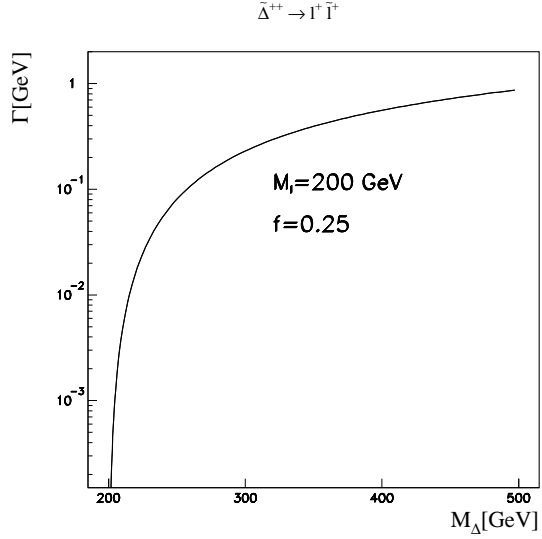} 
%\hfill
%\epsfxsize = 0.5\textwidth \epsffile{dgam2.ps}
}
%\hbox to\textwidth{\hss\epsfig{file=dpl.ps,height=7cm}\hss}
%\vskip -1.cm
%\end{center}
\caption{\it The two--body decay width of  $\Dep$ and $\dep$ versus mass.} 
\label{fig:gam2}
\end{figure}

Since the doubly charged higgsinos are pure states and do not mix 
with other supersymmetric particles,    their decays are 
given by  interactions  not affected by mixing. 
The possible two--body decays of $\Dep$ are 
\bea
\Dep(p) &\ra& \sle^+_L(p_1)\, l^+(p_2)  
\label{Decay2} 
\eea
and
\bea
\Dep(p) &\ra& \tilde{\Delta}^+(p_1)\, W^+(p_2)  
\label{Decay2w}
\eea 
The first decay mode  to a lepton--slepton pair is due to the 
Yukawa interaction \rfn{yuk},  the second decay mode to a singly charged
component
 of 
the triplet and the $W$ boson is due to the weak gauge interactions of the
isotriplet.
Because $\dep$ is an isosinglet, the  only possible two--body decay mode is
given by   
\bea
\dep(p) &\ra& \sle^+_R(p_1) \,l^+(p_2)  
\label{decay2} 
\eea 
Experimental constraints on the $\rho$ parameter imply that 
 members of the same triplet should have masses close to
each other. Therefore the decay \rfn{Decay2w}, with  two 
heavy particles in the final state, is most likely   
not allowed kinematically. In the following we assume that the only
allowed two--body decay for $\Dep$ is the slepton--lepton decay 
mode\footnote{After finalizing this report, a phenomenological analysis of the 
production and GMSB motivated $\tau$ decays of doubly charged higgsinos
at the Tevatron has been presented in Ref.\cite{MTev}.}.
The two--body decay widths are given by
\bea
\Gamma=\frac{f_{\tilde i}^2}{8\pi}
\frac{(M^2_{\tilde i}-m^2_{\tilde{l}})^2}{M^3_{\tilde i}}
\eea
for $\tilde i = \Dep$ and $\dep$. As shown in Fig.\ref{fig:gam2}, 
the widths are small for couplings $f$ of the size of the 
electromagnetic coupling. Denoting the angle between
the polarization vector of the decaying higgsino by 
$\Theta$, the angular distribution is trivially given by
\bea
\frac{1}{\Gamma}\frac{d\Gamma}{d\cos\Theta}=\frac{3}{8}(1\pm\cos\Theta)^2
\eea
for left-- and right--chiral $\tilde\Delta^{++}_L $
and $\tilde\delta^{++}_R $
decays since the scalar couplings flip the chiralities of the massless
leptons.

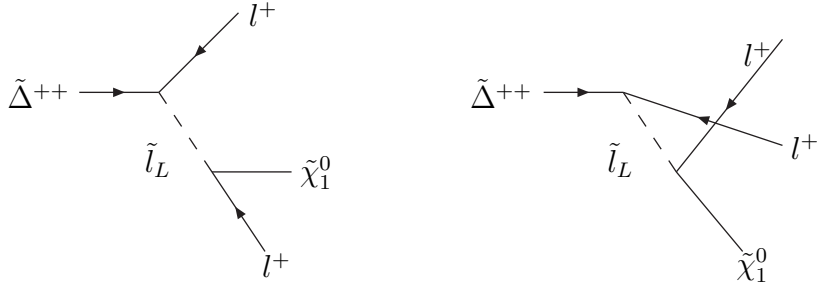
\begin{figure}[t]
\begin{center}
\begin{picture}(350,155)(25,30)
% 1st Diagram
\Text(70,120)[]{$\Dep$}
\ArrowLine(85,120)(115,120)
\ArrowLine(145,150)(115,120)
\Text(155,150)[]{$l^+$}
\DashLine(135,90)(115,120){5}
\Text(115,95)[]{$\sle_L$}
\Line(135,90)(165,90)
%\Photon(185,90)(215,90){3}{7}
\Text(175,90)[]{$\tilde{\chi}^0_1$}
\ArrowLine(155,60)(135,90)
\Text(160,55)[]{$l^+$}
% 2nd Diagram
\Text(245,120)[]{$\Dep$}
\ArrowLine(260,120)(290,120)
\ArrowLine(350,100)(290,120)
\Text(360,100)[]{$l^+$}
\DashLine(290,120)(310,90){5}
\Text(290,95)[]{$\sle_L$}
\ArrowLine(350,140)(310,90)
%\Photon(310,90)(340,90){3}{7}
\Text(342,135)[]{$l^+$}
\Line(335,60)(310,90)
\Text(340,55)[]{$\tilde{\chi}^0_1$}
\end{picture}\\
\end{center}
\caption{\it Diagrams contributing to the three--body decay of $\Dep.$
Analogous diagrams with the intermediate $\sle_R$ give rise to the decay
of
$\dep.$}
\label{fig:decay}
\end{figure}

The two--body decays \rfn{Decay2} and \rfn{decay2} are kinematically 
allowed provided $M_{\Dep},$ $M_{\dep}\gsim m_{\sle}.$ However, 
$\Dep$ and $\dep$ can be 
very light, \ie,  lighter than sleptons. 
In this case, three-body decays of the doubly
charged higgsinos will occur. The possible decay channels are 
\bea
\Dep(p) &\ra& \tilde{\chi}^0(p_1) \,l^+(p_2)\,l^+(p_3) 
\label{Decay3} \\
\Dep(p) &\ra& \tilde{\chi}^+(p_1) \,l^+(p_2)\,\nu(p_3)  
\label{Decay3x} \\
\Dep(p) &\ra& \tilde{\Delta}^+(p_1)\, l^+(p_2)\,\nu(p_3)  
\label{Decay3w}
\eea 
for $\Dep$, and 
\bea
\dep(p) &\ra &\tilde{\chi}^0(p_1)\, l^+(p_2)\,l^+(p_3) 
\label{decay3} \\
\dep(p) &\ra& \tilde{\chi}^+(p_1)\, l^+(p_2)\,\nu(p_3)  
\label{decay3x}
\eea 
for $\dep.$ The decays \rfn{Decay3}, \rfn{Decay3x} and the decays
\rfn{decay3}, \rfn{decay3x} are mediated by virtual left- and 
right-sleptons, respectively, while the decay \rfn{Decay3w} 
is induced by the $W$ boson exchange. 
Assuming the  neutralino $\tilde{\chi}^0_1$ to be the
lightest supersymmetric particle,  
the kinematically  most 
favorable decay modes are \rfn{Decay3} and \rfn{decay3}. 
In the following analysis we assume  that the
three--body decays  \rfn{Decay3} and \rfn{decay3}, 
cf. Fig.\ref{fig:decay}, are the only modes which are allowed kinematically. 

The matrix elements of the three-body decays (\ref{Decay3},\ref{decay3}) 
consist of two terms corresponding to the two diagrams
in Fig.\ref{fig:decay}.  These can be expressed as
\begin{eqnarray}
{\cal D}_{L,R}
& =&\frac{4ef}{\sqrt{2}}\bigg[
F^1_{L,R}\left[\bar{u}(l^+_1)\,P_{L,R}\, u(++)\right]
         \left[\bar{u}(\x)\,P_{L,R}\, v(l^+_2)\right] \nn \\
& +&
F^2_{L,R}\left[\bar{u}(l^+_2)\,P_{L,R}\, u(++)\right]
         \left[\bar{u}(\x)\,P_{L,R}\, v(l^+_1)\right]
\bigg]
\label{3amp}
\end{eqnarray}
with the form factors
\begin{eqnarray}
F^1_L=\frac{\cot 2\theta_W N_{12}^* + N_{11}^*}
             {s_1-m^2_{\sle_L}+im_{\sle_L}\Gamma_{\sle_L}}\,,
& \,\,\,\,&
F^2_L=\frac{\cot 2\theta_W N_{12}^* + N_{11}^*}
             {s_3-m^2_{\sle_L}+im_{\sle_L}\Gamma_{\sle_L}} 
 \nonumber\\
F^1_R=\frac{\tan\theta_W N_{12} - N_{11}}
             {s_1-m^2_{\sle_R}+im_{\sle_R}\Gamma_{\sle_R}}\,, 
& \,\,\,\,&
F^2_R=\frac{\tan\theta_W N_{12} - N_{11}}
             {s_3-m^2_{\sle_R}+im_{\sle_R}\Gamma_{\sle_R}}  
\label{efs}
\end{eqnarray}
The terms denoted with $L$ and $R$ correspond to the decays of
$\Dep$ and $\dep,$ respectively, and $N_{11},$ $N_{12}$ are the elements of
the unitary matrix diagonalizing the neutralino mass matrix in the basis
$\tilde{\gamma},$ $\tilde{Z},$ $\tilde{H}^0_a,$ $\tilde{H}^0_b$ \cite{mssm}.  
The Mandelstam variables $s_1$, $s_2$, $s_3$ in the form factors
are defined in terms of the 4--momenta of the final state particles as
$s_1=(p_1+p_2)^2,  s_2=(p_2+p_3)^2$ and $s_3=(p_1+p_3)^2\,.$ 
\begin{figure}[t]
%\begin{center}
\centerline{
\epsfxsize = 0.5\textwidth \epsffile{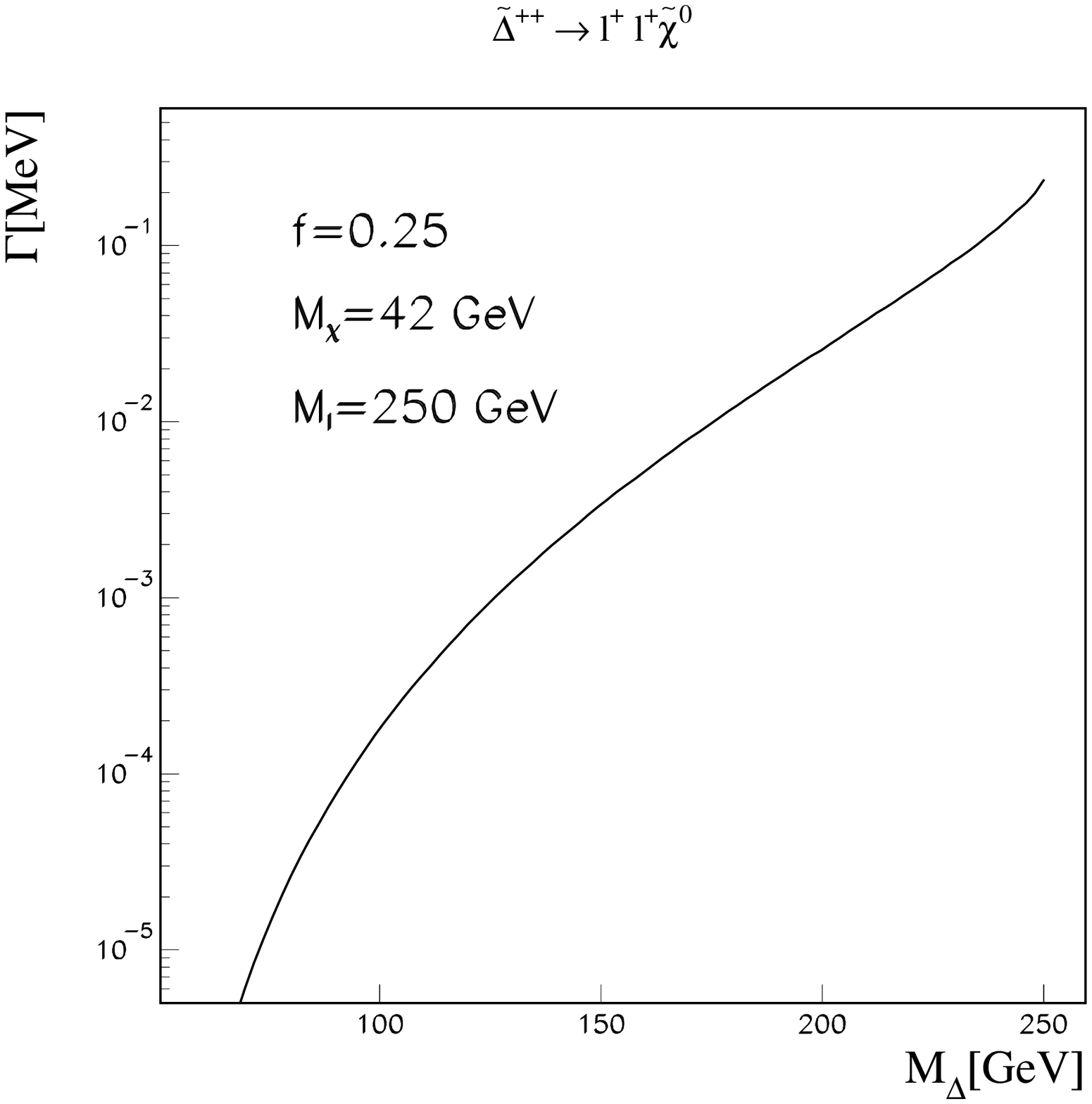} \hfill
\epsfxsize = 0.5\textwidth \epsffile{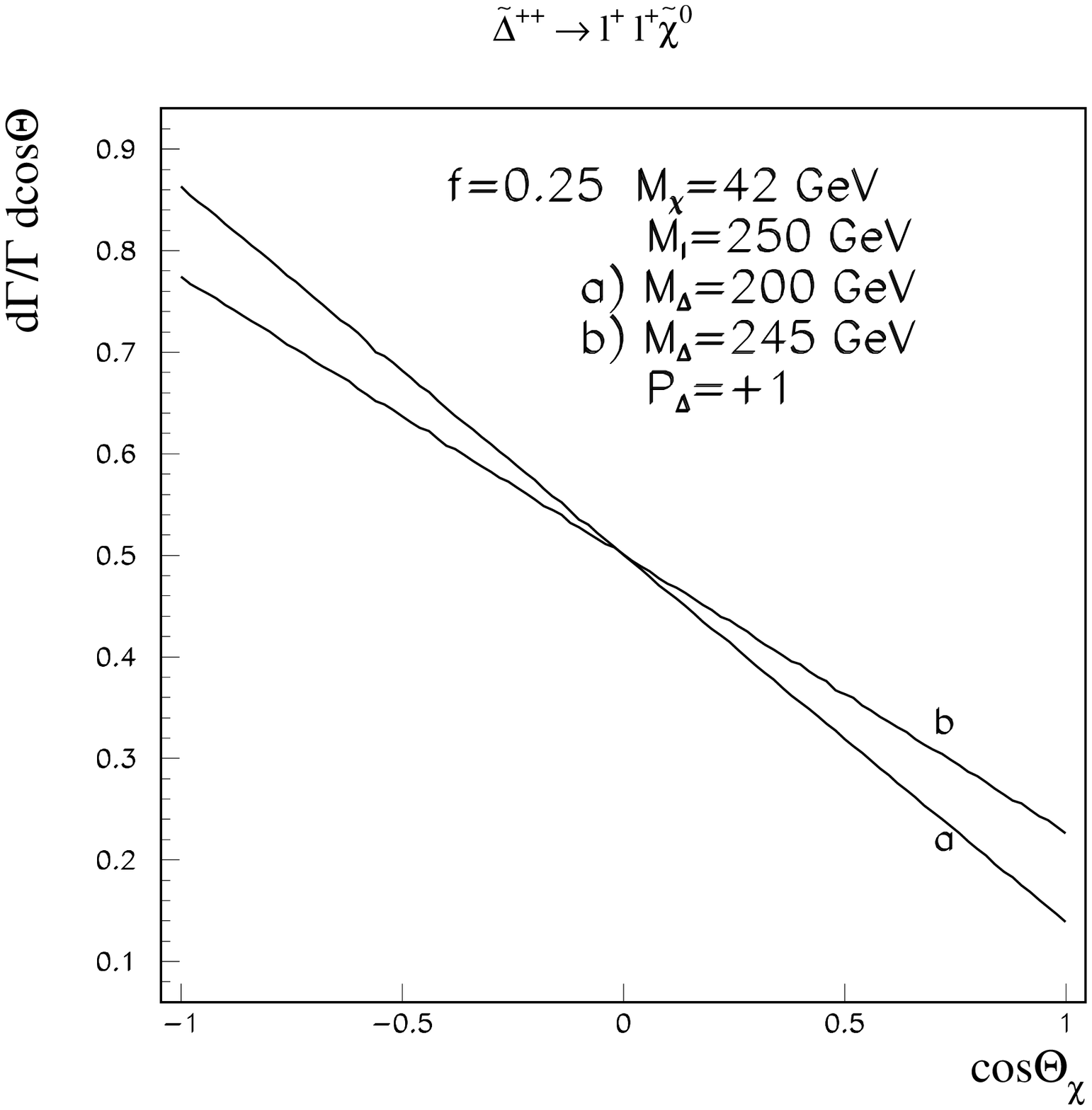}
}
%\hbox to\textwidth{\hss\epsfig{file=dpl.ps,height=7cm}\hss}
%\vskip -1.cm
%\end{center}
\caption{\it The three-body decay width of  $\Dep$ 
and  the normalized angular distributions of 
the final state $\x$ in the rest frame of the polarized $\Dep.$
Numerical values for the parameters are indicated in the figures.}
\label{fig:gam3}
\end{figure}
The decay widths and distributions 
of $\Dep,$ $\dep$ with  polarization vector $\vec{\cal P}$ 
can be found from the following expression
\begin{eqnarray}
\frac{d\Gamma (n)}{d PS}  &=& \frac{4\pi \alpha f}{ M} \bigg\{
-(s_1- M^2)(s_1- m_{\x}^2) |F^1_{L,R}|^2 
-(s_3- M^2)(s_3- m_{\x}^2) |F^2_{L,R}|^2 \nonumber\\
&& +2 (s_1s_3 -M^2 m^2_{\x})  {\rm Re}(F^1_{L,R} F^{2*}_{L,R})
    \nonumber \\
&& +\eta_{L,R} 2 (n\cdot p_2) M [
  (s_1-M^2){\rm Re}(F^1_{L,R} F^{2*}_{L,R})
-(s_3 - m_{\x}^2) |F^2_{L,R}|^2] \nonumber\\
&& +\eta_{L,R} 2(n\cdot p_3) M [
  (s_3-M^2){\rm Re}(F^1_{L,R} F^{2*}_{L,R})
-(s_1 - m_{\x}^2) |F^1_{L,R}|^2]  \bigg\}
\label{3d}
\end{eqnarray}
where $n_\mu$ is the $\Dep$($\dep$) spin 4--vector and
$dPS$ the phase--space element,
\bea
dPS=
\frac{1}{(2\pi)^5}\frac{1}{32 M^2}
ds_1 ds_2 d\Omega_1^* d\phi^*_{ll}
\eea
 $\Omega_1^*=(\theta^*,\phi^*)$
describes  orientation of the neutralino in the rest frame of the  
doubly charged higgsino and $\phi^*_{ll}$ measures the rotation of 
the recoiling dilepton  
system about this axis. 
Again,  $L$ and $R$ correspond to the decay of $\Dep$ and
$\dep,$ 
respectively, and $\eta_L=1,$ $\eta_R=-1.$
In Fig.\ref{fig:gam3} 
 the three--body decay width for $\Dep$ is shown as a 
 function of the mass. 
In the numerical example,
the neutralino 
mixing components  are chosen as  $N_{11}=0.94$ and 
$N_{12}=-0.32,$ corresponding to the MSSM parameters $\tan\beta=2,$
$M_2=78$ GeV  and $\mu=-250$ GeV. For these parameters the mass of the
lightest neutralino is $m_{\x_1}=42$ GeV \cite{R6A}.
Because the amplitude \rfn{3amp}
depends linearly on the combination of $N_{ij}$, cf.
 \eq{efs},
 the dynamics of the three--body decays \rfn{Decay3}, \rfn{decay3}
does not depend on the choice of these parameters: 
the distributions are the same for  different parameter sets. 
The neutralino  mixing parameters are assumed to be 
known from other collider experiments.  

If the angular distribution in the $l^+l^+$ rest system is integrated
out, the $\Dep$ and $\dep$ three--body 
decay final states are described by the energy 
and the polar angle of $\tilde{\chi}^0_1,$ or equivalently 
by the energy and the
polar angle of the $l^+l^-$ pair, which can be measured directly.  
In Fig.\ref{fig:gam3}  the normalized 
angular distribution of the final state $\x$
in polarized $\Dep$ decays are illustrated 
in the rest frame of $\Dep.$ The distributions
depend on the masses of the particles and they are opposite for different 
polarization states. For the same set of masses 
the distributions in  the polarized decays of $\dep$ 
are identical to  $\Dep$ yet with the opposite sign.

For the subsequent analysis of the angular correlations between the two 
doubly charged higgsinos  in the processes \rfn{DD}, \rfn{dd}, 
it is convenient to determine the normalized spin--density 
matrix elements $\rho_{\lambda\lambda'}\sim {\cal D}_\lambda {\cal 
D}^*_{\lambda'}$
for the kinematical configuration described above. Choosing 
the $\Dep$ flight direction as quantization axis, the 
spin--density matrices are  given by the expressions 
\begin{eqnarray}
&& \rho_{\lambda\lambda^\prime}
   =\frac{1}{2}\left(\begin{array}{cc}
         1+\kappa\cos\theta^*   &   \kappa\sin\theta^*{\rm e}^{i\phi^*} \\
 \kappa\sin\theta^*{\rm e}^{-i\phi^*} &        1-\kappa\cos\theta^*
                     \end{array}\right)\nonumber\\
&& \bar{\rho}_{\bar{\lambda}\bar{\lambda}^\prime}
   =\frac{1}{2}\left(\begin{array}{cc}
 1+\bar{\kappa}\cos\bar{\theta}^* & 
  \bar{\kappa}\sin\bar{\theta}^*{\rm e}^{i\bar{\phi}^*} \\
  \bar{\kappa}\sin\bar{\theta}^*{\rm e}^{-i\bar{\phi}^*} & 
 1-\bar{\kappa}\cos\bar{\theta}^* \end{array}\right)
\label{3rho}
\end{eqnarray}
$\theta^*$ ($\bar{\theta}^*$) is the polar angle of the 
$l^+l^+$ ($l^- l^-$) system in the 
$\Dep$ ($\Dem$) rest frame with respect to the original 
flight direction in the laboratory frame, and $\phi^*$ ($\bar{\phi}^*$) 
is the 
corresponding azimuthal angle with respect to the production plane. 
The spin analysis--power $\kappa$, which measures the left--right
asymmetry,
depends on the particle masses and couplings involved in the decay,  and
on 
the Mandelstam variable $s_2$ which is a square of the invariant mass
of the final--state $l^+l^+$ system. 
 Neglecting  small effects from
 non--zero widths, loops and CP--noninvariant phases, $\kappa$ (and 
$\bar{\kappa}$) is real. 
\begin{figure}[t]
%\begin{center}
\centerline{
\epsfxsize = 0.5\textwidth \epsffile{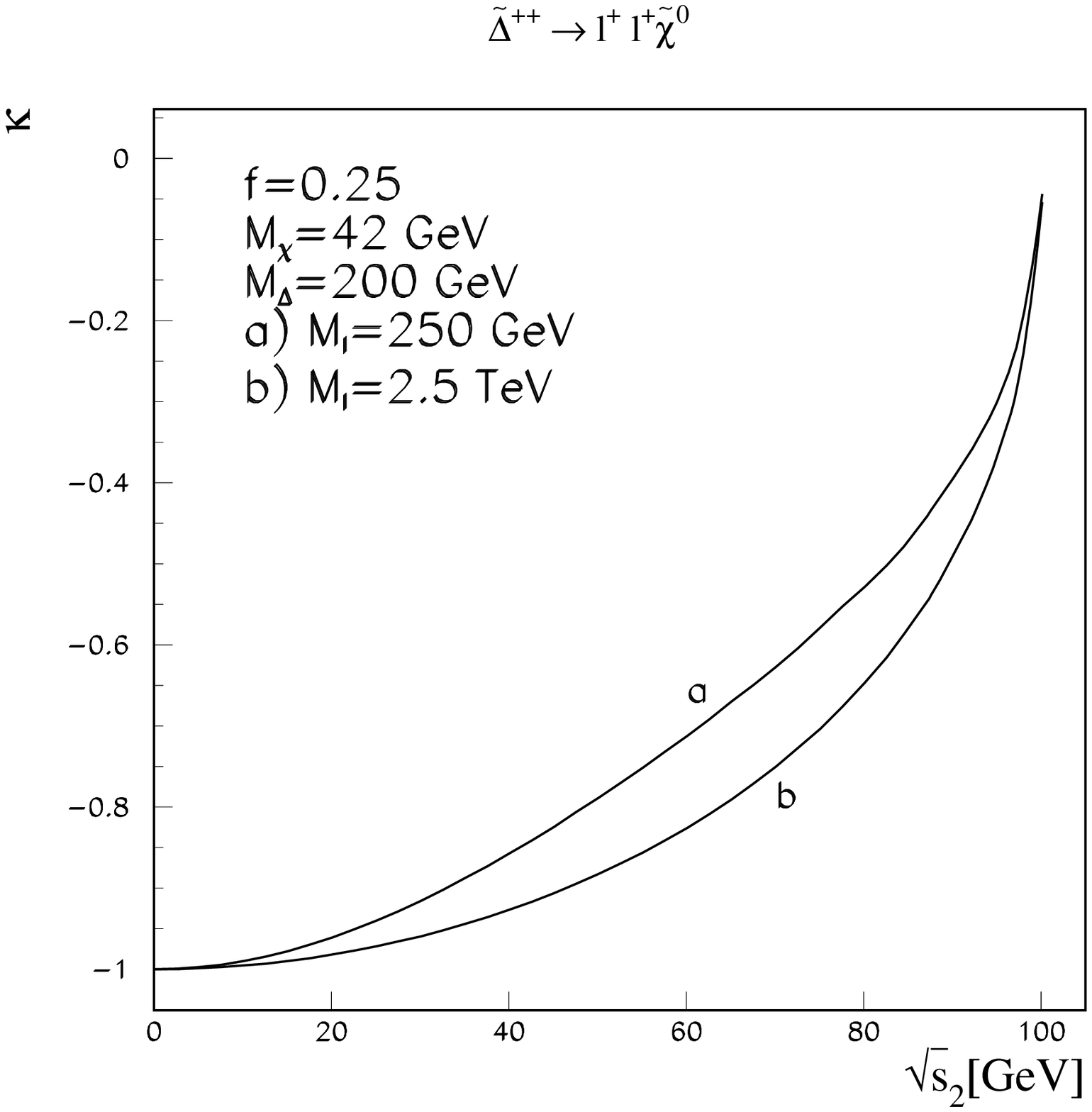} 
\hfill
\epsfxsize = 0.5\textwidth \epsffile{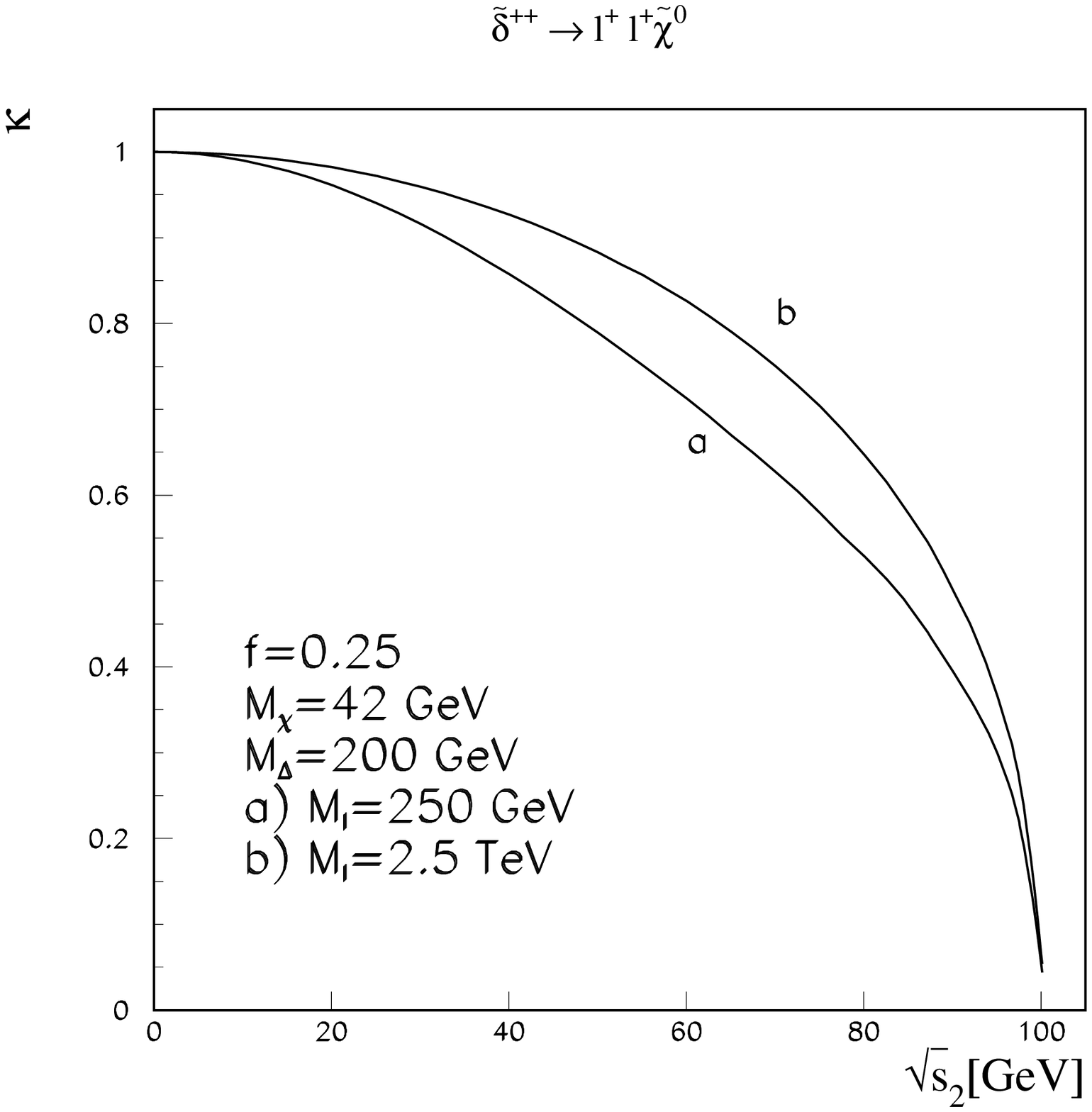}
}
%\hbox to\textwidth{\hss\epsfig{file=dpl.ps,height=7cm}\hss}
%\vskip -1.cm
%\end{center}
\caption{\it Values of the paramater $\kappa$ in the decays \rfn{Decay3}
and \rfn{decay3} as functions of  
the invariant mass of the final state $l^+l^+$
system
for the masses indicated in the figure. 
}
\label{fig:kap}
\end{figure}
The paramater $\kappa$ in 
the decays \rfn{Decay3} and \rfn{decay3} is shown in Fig.\ref{fig:kap}
as a function of the invariant mass $\sqrt{s_2}$ 
of the final state $l^+l^+$ system. 
 $\kappa$ is in general large 
over the whole range of the invariant mass and approaches
zero only at the kinematical limit. If the virtual sleptons mediating 
the decay are very heavy (curve $b$ in Fig.\ref{fig:kap}), 
 the slepton propagators can be approximated
by point propagators. In this case, the analytic expression for $\kappa$
in the $\Dep$ decay  is particularly simple, 
\begin{eqnarray}
\kappa(s_2)=\frac{\lambda \left[2(M_{\Dep}^2-m_{\x}^2)-s_2\right]}
{\lambda^2+3 s_2(M_{\Dep}^2+m_{\x}^2)-3(M_{\Dep}^2-m_{\x}^2)^2}     
\end{eqnarray}
where   $\lambda = [M_{\Dep}^2-(\sqrt{s_2}-m_{\x}^2)^2]^{1/2} 
[M_{\Dep}^2-(\sqrt{s_2}+m_{\x}^2)^2]^{1/2}.$
For the decay \rfn{decay3} of $\dep$ the parameter $\kappa$ 
has the opposite sign but  the  
absolute value is numerically equal to the  $\Dep$ decay.

%%%%%%%%%%%%%%%%%%%%%%%%%%%%%%%%%%%%%%%%%%%%%%%%%%
\subsubsection*{4.2 Angular Correlations}
\label{sec:angular correlation}
%%%%%%%%%%%%%%%%%%%%%%%%%%%%%%%%%%%%%%%%%%%%%%%%%%

The two--fold ambiguity in extracting the  $\Dep$ and $\dep$ 
couplings from the total cross sections can be resolved
by measurements of angular correlations which reflect
the polarization states of the higgsinos in the production
processes:
\begin{center}
\begin{picture}(300,100)(0,0)
\Text(130,80)[]{$e^+e^-\rightarrow\Dem\Dep$}
\Line(132,70)(132,40)
\Line(155,70)(155,60)
\Line(132,40)(162,40)
\Text(155,61)[l]{$\rightarrow\x +(l^+l^+)$}
\Text(147,40)[l]{$\longrightarrow\hskip 0.0cm\x+(l^-l^-)$}
\end{picture}
\end{center}
\vskip -1cm
The analyses are complicated since the two invisible neutralinos in the 
final state
do not allow for a complete reconstruction of the events. In particular,
it is not possible to measure the $\Dep, \dep$ production angle
$\Theta$; this angle can be determined only up to a two--fold ambiguity
which, however, becomes increasingly less effective with
rising energy.

The final state distributions can be  found by combining 
the polarized cross section with the polarized decay distributions.
After integrating over the  unobservable  production angle
$\Theta$ and  
the $(l^+l^+)$
and $(l^-l^-)$ invariant masses, the integrated cross section
\begin{eqnarray}
&&\frac{{\rm d}^4\sigma\left[e^+e^-\rightarrow\Dep\Dem
               \rightarrow \tilde{\chi}^0_1\tilde{\chi}^0_1(l^+l^+)
               (l^-l^-)\right]}{{\rm d}\cos\theta^*{\rm d}\phi^*
               {\rm d}\cos\bar{\theta}^*{\rm d}\bar{\phi}^*} \hskip 1.0cm
\nonumber\\
&& \hskip 0.4cm =\frac{\alpha^2\beta}{128\pi s} 
   {\rm Br}(\Dep\rightarrow\tilde{\chi}^0_1 l^+l^+)
   {\rm Br}(\Dem\rightarrow\tilde{\chi}^0_1 l^-l^-)
    \Sigma(\theta^*,\phi^*;\bar{\theta}^*,\bar{\phi}^*)
\;\;\;\;\;\;\;\;\;\;\;\;\;\;\;\;\;\;\;\;\;
\end{eqnarray}
can be decomposed into sixteen independent angular parts.
In addition to the unpolarized cross section, four angular
distributions can be determined experimentally even though two 
neutralinos escape undetected:
\begin{eqnarray}
\label{sigma}
\Sigma&=&\Sigma_{\rm unpol}+\cos\theta^*\kappa{\cal P}
       +\cos\bar{\theta}^* \bar{\kappa}\bar{\cal P}
       +\cos\theta^*\cos\bar{\theta}^*\kappa\bar{\kappa}{\cal
Q}\nonumber\\
     &&+\sin\theta^*\sin\bar{\theta}^*\cos(\phi^*+\bar{\phi}^*)
        \kappa\bar{\kappa}{\cal Y} + .....
\end{eqnarray}
The ellipsis denotes the remaining orthogonal angular distributions 
which cannot be measured.
The polarizations ${\cal P}$ and $\bar{\cal P}=-{\cal P}$ have been expressed
in terms of the generalized charges in \eq{PL}. Analogously, 
\begin{figure}[t]
%\begin{center}
\centerline{
\epsfxsize = 0.5\textwidth \epsffile{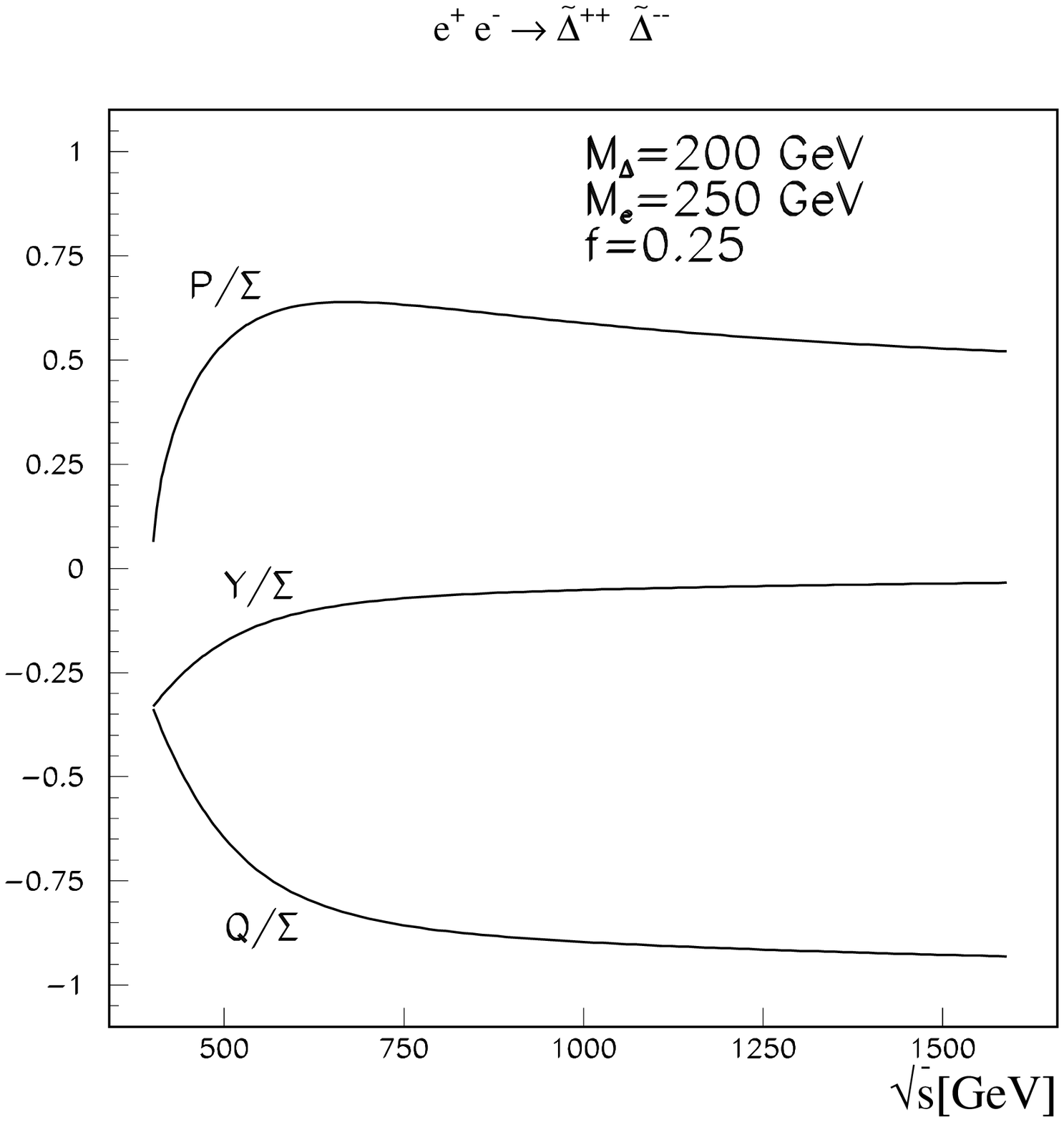}
\hfill
\epsfxsize = 0.5\textwidth \epsffile{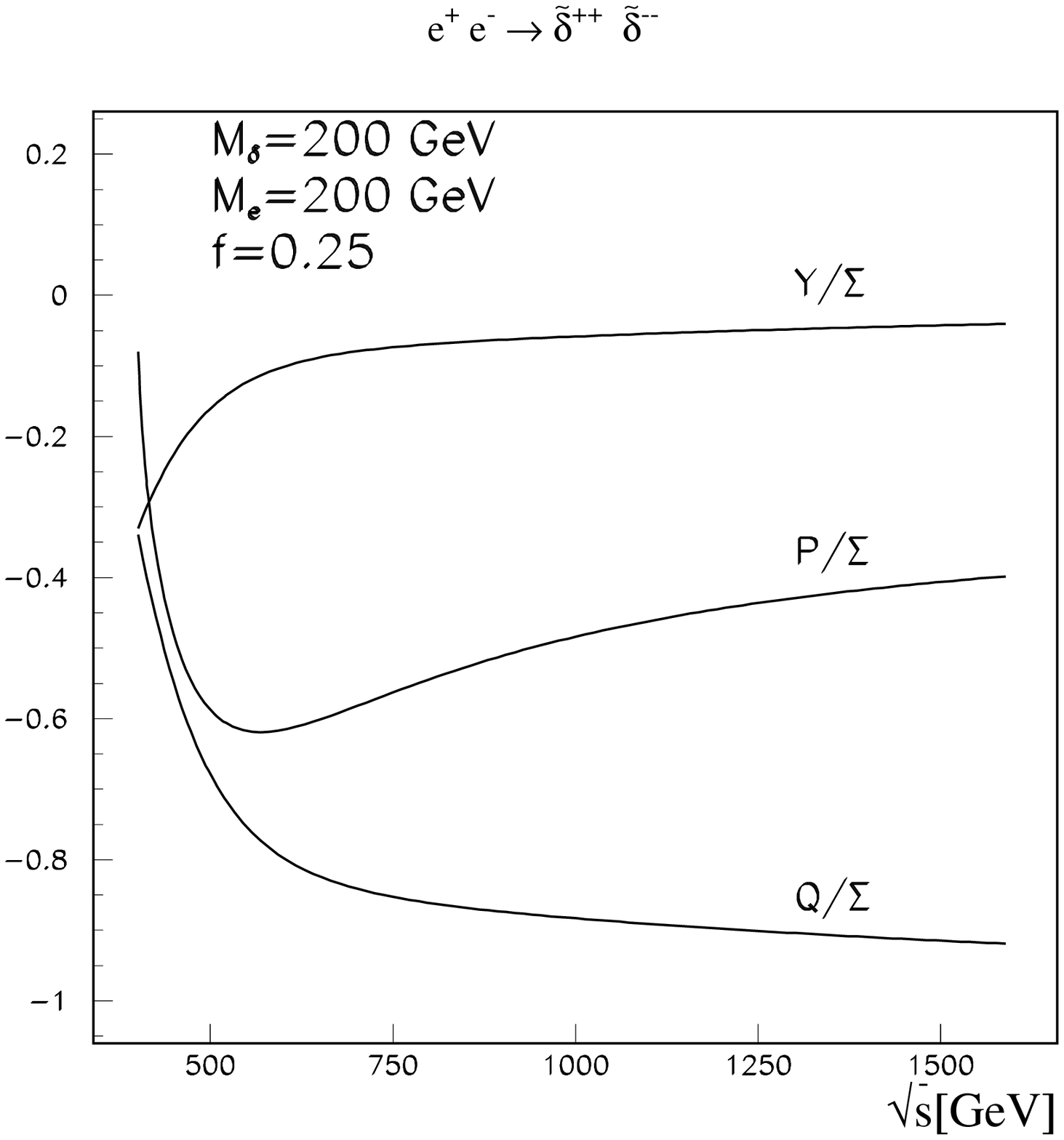}
}
%\hbox to\textwidth{\hss\epsfig{file=dpl.ps,height=7cm}\hss}
%\vskip -1.cm
%\end{center}
\caption{\it Correlation functions ${\cal P},$ ${\cal Q}$ and ${\cal Y}$
normalized to $\Sigma_{\rm unpol}$
as functions of the collision energy for the processes $\ee\ra\Dep\Dem$
and $\ee\ra\dep\dem$.}
\label{fig:pqy}
\end{figure}
 the correlation functions 
${\cal Q}$ and ${\cal Y}$ are defined by the charges in the following
way:
\begin{eqnarray}
&& {\cal Q}=-4\int {\rm d}\cos\Theta\left[(\beta^2+\cos^2\Theta)Q_1
         +(1-\beta^2)\cos^2\Theta Q_2+2\beta\cos\Theta Q_3\right]\nonumber
\\
% && {\cal Z}=-2\int {\rm d}\cos\Theta\left[(1-\beta^2)Q_1
%          +(1+\beta^2)Q_2\right]\sin^2\Theta 
&& {\cal Y}=-2\int {\rm d}\cos\Theta(1-\beta^2)\left[Q_1+Q_2\right]
\sin^2\Theta 
\end{eqnarray}
The  energy dependence of the three correlation functions, 
${\cal P},$ ${\cal Q}$ and  ${\cal Y,}$  normalized  to 
$\Sigma_{\rm unpol}$ is shown in Fig.\ref{fig:pqy}. 
For the both processes $\ee\ra\Dep\Dem$ and $\ee\ra\dep\dem$
they are smooth functions of the c.m. energy, quite 
 different from zero for all values of the 
 energy.

%%%%%%%%%%%%%%%%%%%%%%%%%%%%%%%%%%%%%%%%%%%%%%%%%%
\subsection*{5. Observables  of the Doubly Charged Higgsinos} 
\label{sec:observable}
%%%%%%%%%%%%%%%%%%%%%%%%%%%%%%%%%%%%%%%%%%%%%%%%%%

%%%%%%%%%%%%%%%%%%%%%%%%%%%%%%%%%%%%%%%%%%%%%%%%%%
\subsubsection*{5.1 Signals and Background}
%%%%%%%%%%%%%%%%%%%%%%%%%%%%%%%%%%%%%%%%%%%%%%%%%%

The final states of the doubly charged higgsino pair production
processes \rfn{DD}, \rfn{dd} and  their subsequent decays \rfn{Decay2},
\rfn{decay2} or \rfn{Decay3}, \rfn{decay3} consist of four charged
leptons $l_1^+l_1^+l_2^-l_2^-$ plus missing energy $\not\!\!E$. 
Because $\tilde\Delta^{++}$, $\tilde\delta^{++}$ 
carry two units of lepton number and two units of
electric charge, the final state leptons with the same charge and flavor
are associated with the same decaying particle:
\begin{eqnarray}
\ee & \ra & \Dep \Dem,\; \dep \dem \nonumber \\
& \ra & (l^+l^+)(l^-l^-) + \!\!\not\!\!E
\end{eqnarray}
The final--state observables of the signals are therefore clearly
distinct from the background processes in which same--sign 
dileptons originate necessarily from decays of different particles.

Indeed, the main SM background is due to triple gauge--boson production:
$\ee \ra \gamma^* W^+W^- \ra l^+l^- l^+l^- +\!\!\not\!\!E $ etc. These
processes are of higher order in the gauge couplings so that the cross 
sections are small and the backgrounds are under control. SUSY--induced
backgrounds are generated by the production of
heavy neutralinos that can decay into  $\tilde\chi^0_1$ and a 
charged lepton pair: $\ee \ra \tilde{\chi}^0_i  \tilde{\chi}^0_j
\ra l^+l^-l^+l^- +\!\!\not\!\!E $. Since, in both cases, the kinematical
configurations of the background and signal events are 
very different from each other, the large number
of  events can be used to enrich the sample of 
signal events statistically by applying suitable cuts
 to disentangle signal from background effects.

\underline{{\bf Mass}}. The masses $M_{\tilde\Delta^{++}}$ and
 $M_{\tilde\delta^{++}}$
can be measured very precisely near the thresholds where the production
cross sections $\sigma (\ee \ra \tilde{\Delta}^{++} 
\tilde{\Delta}^{--} / \tilde{\delta}^{++} 
\tilde{\delta}^{--} )$ rise sharply with the 
velocities $\beta = \sqrt{1-4 M^2_{\tilde{\Delta}^{++}} / s}$
and $\sqrt{1-4 M^2_{\tilde{\delta}^{++}} / s}$,
respectively. With beam parameters as anticipated
for TESLA, a precision better than $\sim$ 100 MeV can
be achieved by this method for an  integrated luminosity of 50 fb$^{-1}$
\cite{R4}.

\underline{{\bf Isospin}}. 
As evident from eq.(\ref{charges}), 
the isospin of
the two states $\tilde{\Delta}^{++}$ and 
$\tilde\delta^{++}$ can be measured by using 
right-- and left--handedly polarized electron beams. The two 
cross sections depend quadratically on the isospin,
\bea
\sigma (\ee \ra \tilde{\Delta}^{++} \tilde{\Delta}^{--}) =
\frac{4 \pi \alpha^2}{s} \beta \left[ \, 1 + c(s) I_3  \, \right]^2
\eea
yet the root ambiguity can easily be resolved by carrying out
the measurements at two different beam energies. The vector--character
of the states can be proven by establishing the vanishing of 
forward--backward asymmetries.

\underline{{\bf Trilinear couplings $f$}}. Once the isospin
components are determined, the couplings $f_{\tilde\Delta}$ and
$f_{\tilde\delta}$ can be probed by measuring
the cross sections in the mirror processes for right--
and left--handedly polarized electron
beams, respectively, cf. eqs.(\ref{qD},\ref{qd}). The selectron masses
are assumed to  be known from the pair 
production of these particles. The measurements of the cross 
sections lead, in general, to a two--fold ambiguity in $f$.
This ambiguity can be resolved in two ways. (i) If large enough 
variations in the cm energy are possible, measurements at two
different energy points give rise to two independent
equations for $f$ with only one common solution for both. 
(ii) The ambiguity can also be resolved by analyzing spin
correlations. Since two invisible neutralinos are 
present in the final states after the decays of $\tilde\Delta$
and $\tilde\delta$, the kinematics of the $\tilde\Delta$
and $\tilde\delta$ states cannot be 
reconstructed completely. Nevertheless, the polar angles
$\vartheta_*$ and the product of the transverse momentum
vectors of the two $\tilde\chi^0_1$'s can be expressed
by measured energies and laboratory angles,
\begin{eqnarray}
&&\cos\theta^*=\frac{1}{\beta |\vec{p}^*|}
\left(\frac{E}{\gamma}-E^*\right)
\ \ {\rm and} \ \  \cos\bar{\theta}^*=\frac{1}{\beta |\vec{\bar{p}}^*|}
               \left(\frac{\bar{E}}{\gamma}-\bar{E}^*\right)  \nonumber\\
&&\sin\theta^*\sin\bar{\theta}^*\cos(\phi^*+\bar{\phi}^*)=
         \frac{|\vec{p}||\vec{\bar{p}}|\cos\vartheta+\left(E-E^*/\gamma
\right)
         \left(\bar{E}- \bar{E}^*/\gamma\right)}{\beta^2 
         |\vec{p}^*||\vec{\bar{p}}^*|} 
\end{eqnarray}
where $\gamma=\sqrt{s}/{2M_{\Dep}}$ etc.; $\vartheta$ is the 
angle between the momenta of the two lepton systems with 
opposite charges. 
As evident from \eq{sigma}, the degree of polarizations
 ${\cal P}$ and $\bar{\cal P}$, and the spin correlations
${\cal Q}$ and ${\cal Y}$ can   be measured directly 
despite the two neutralinos  escaping detection.

The correlation functions come with the spin analysis--powers 
$\kappa$ and $\bar{\kappa}$
which depend on masses and couplings involving the neutralinos
$\tilde{\chi}^0_1$. The $\kappa, \bar{\kappa}$ dependence however 
factorizes out of the
correlation functions so that these parameters can be eliminated by taking
appropriate ratios. As a result, two independent observables can be
constructed
from angular correlations, which can be measured directly in terms of 
laboratory momenta: ${\cal P}^2/{\cal Q}$ and ${\cal Y}/{\cal Q}$. 
\begin{figure}[t]
%\begin{center}
\centerline{
\epsfxsize = 0.5\textwidth \epsffile{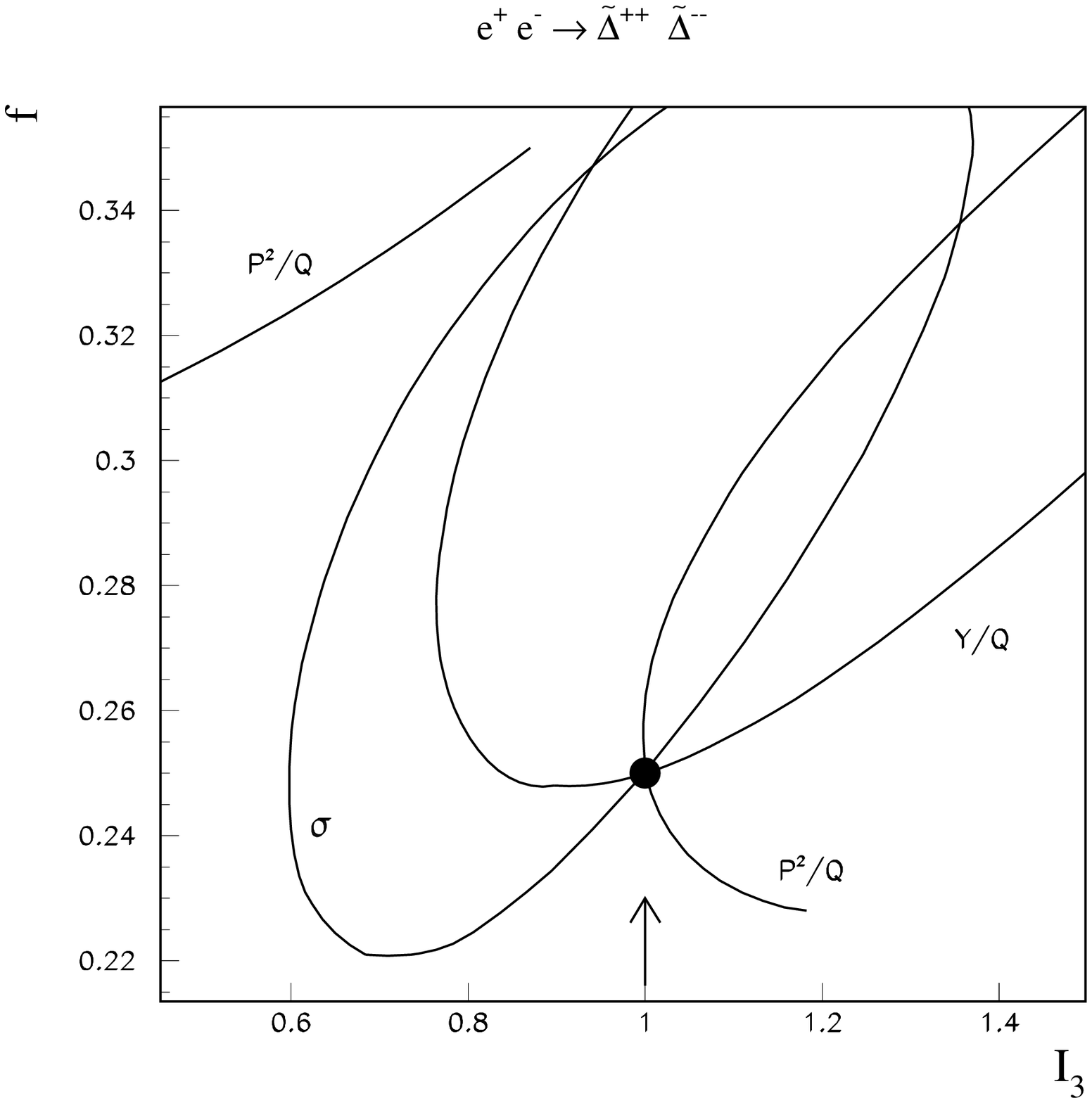}
\hfill
\epsfxsize = 0.5\textwidth \epsffile{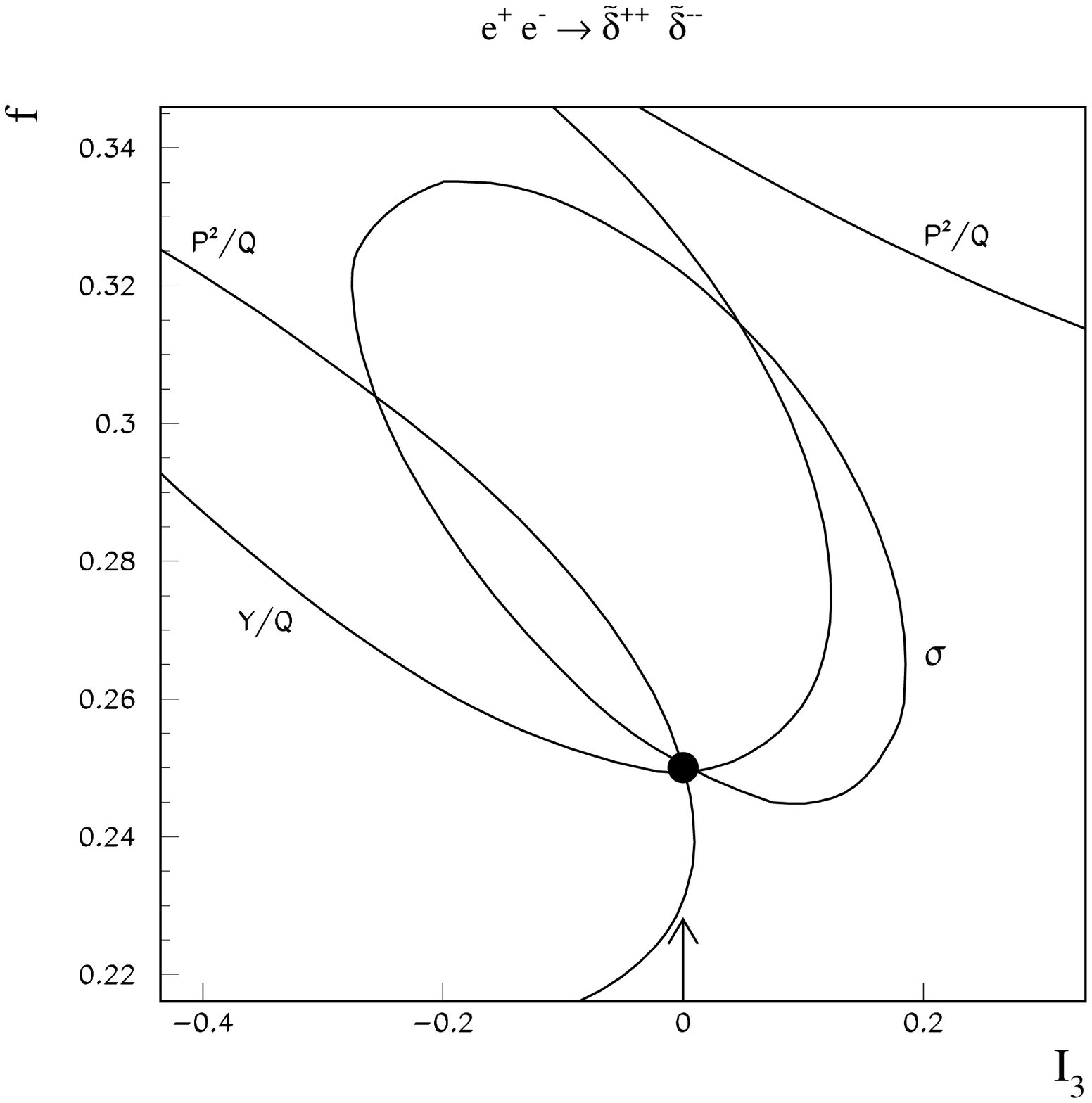}
}
%\hbox to\textwidth{\hss\epsfig{file=dpl.ps,height=7cm}\hss}
%\vskip -1.cm
%\end{center}
\caption{\it Contour lines from the measurements of 
 $\sigma,$  ${\cal P}^2/{\cal Q}$ and ${\cal Y}/{\cal Q}$
determining the couplings $f$ and the weak isospin $I_3$
of the doubly charged higgsinos $\tilde{\Delta}^{++}$ and
$\tilde{\delta}^{++}$. The common crossing point is indicated by the dot.
}
\label{fig:pqycurv2}
\end{figure}

To demonstrate that the ambiguity can be resolved by
measuring the spin correlations, we assume, in a 
{\it Gedanken--experiment}, a set of ``measured observables'',
\ie  unpolarized cross sections and  
correlation ratios; for illustration:
\begin{center}
\begin{tabular}{cccc}
$\Dep$ : & $\sigma$ = 0.139~pb & ${\cal P}^2/{\cal Q}$ = -2.81 &
${\cal Y}/{\cal Q}$ = 0.27 \\
$\dep$ : & $\sigma$ = 0.093~pb & ${\cal P}^2/{\cal Q}$ = -2.12 &
${\cal Y}/{\cal Q}$ = 0.23 \\
\end{tabular}
\end{center}
The contour lines of these observables\footnote{The contour lines are in
general not closed curves, yet may consist of several disconnected branches.} 
are shown in Figs.\ref{fig:pqycurv2} 
in the planes [$I_3, f$], assuming the collision energy to be 
$\sqrt{s}=500$ GeV and higgsino and selectron masses $200$ GeV and 250 GeV,
respectively. While the contours of
$\sigma, {\cal P}^2/{\cal Q}, {\cal Y}/{\cal Q}$ give pairwise rise to
several intersections, the curves cross each other, all at
the same time, only once. Moreover, this crossing point is the
only solution that is consistent with
integer/half--integer values of $I_3$, i.e.
$I_3 (\tilde{\Delta}^{++}) = + 1$ and $I_3 (\tilde{\delta}^{++}) = 0$.
Thus a unique solution for quantum numbers and couplings
can be extracted from the measurements of 
(un)polarized cross sections, forward--backward asymmetries, and
spin--spin correlations.

\begin{figure}[t]
%\begin{center}
\centerline{
\epsfxsize = 0.5\textwidth \epsffile{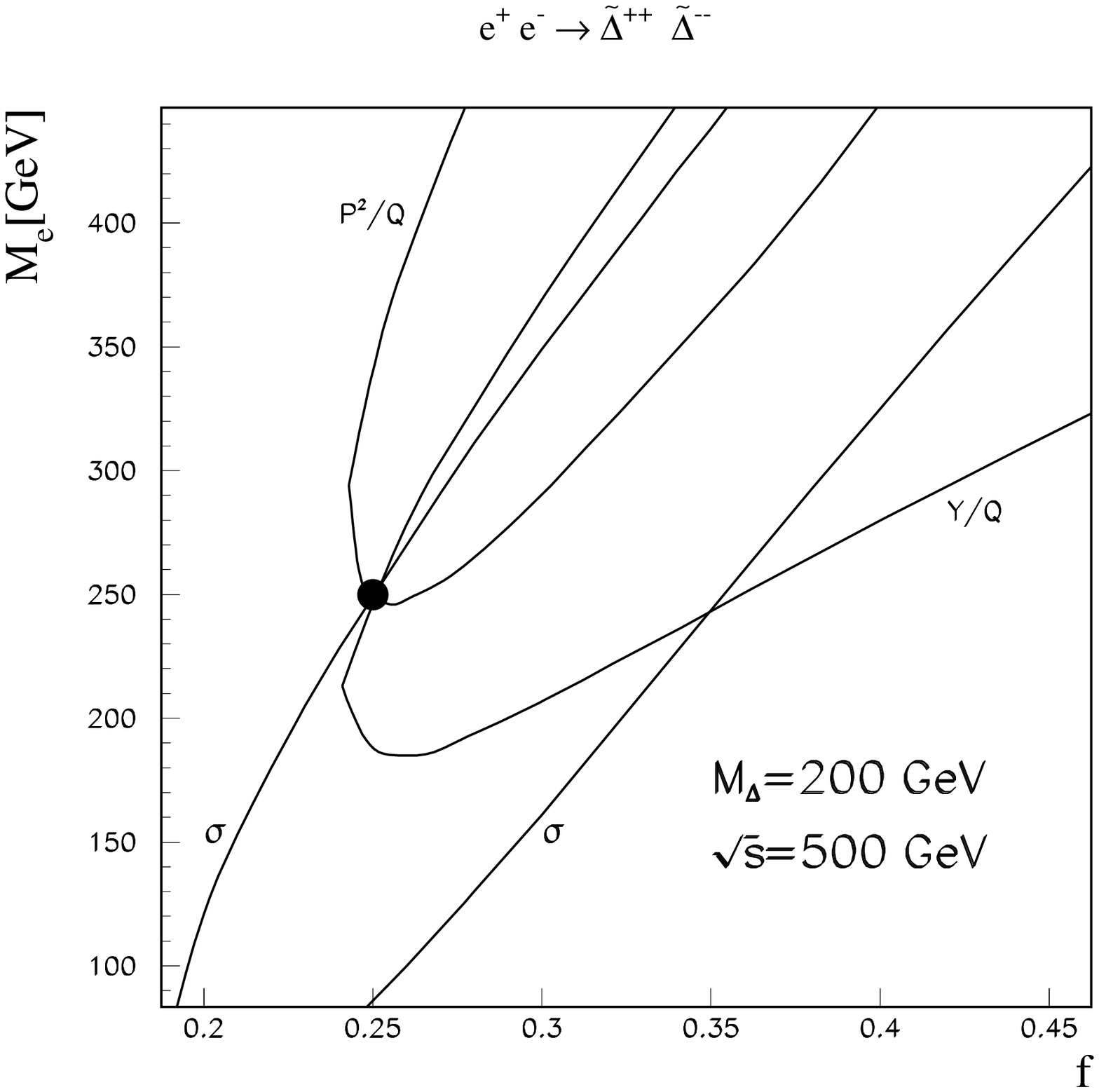}
\hfill
\epsfxsize = 0.5\textwidth \epsffile{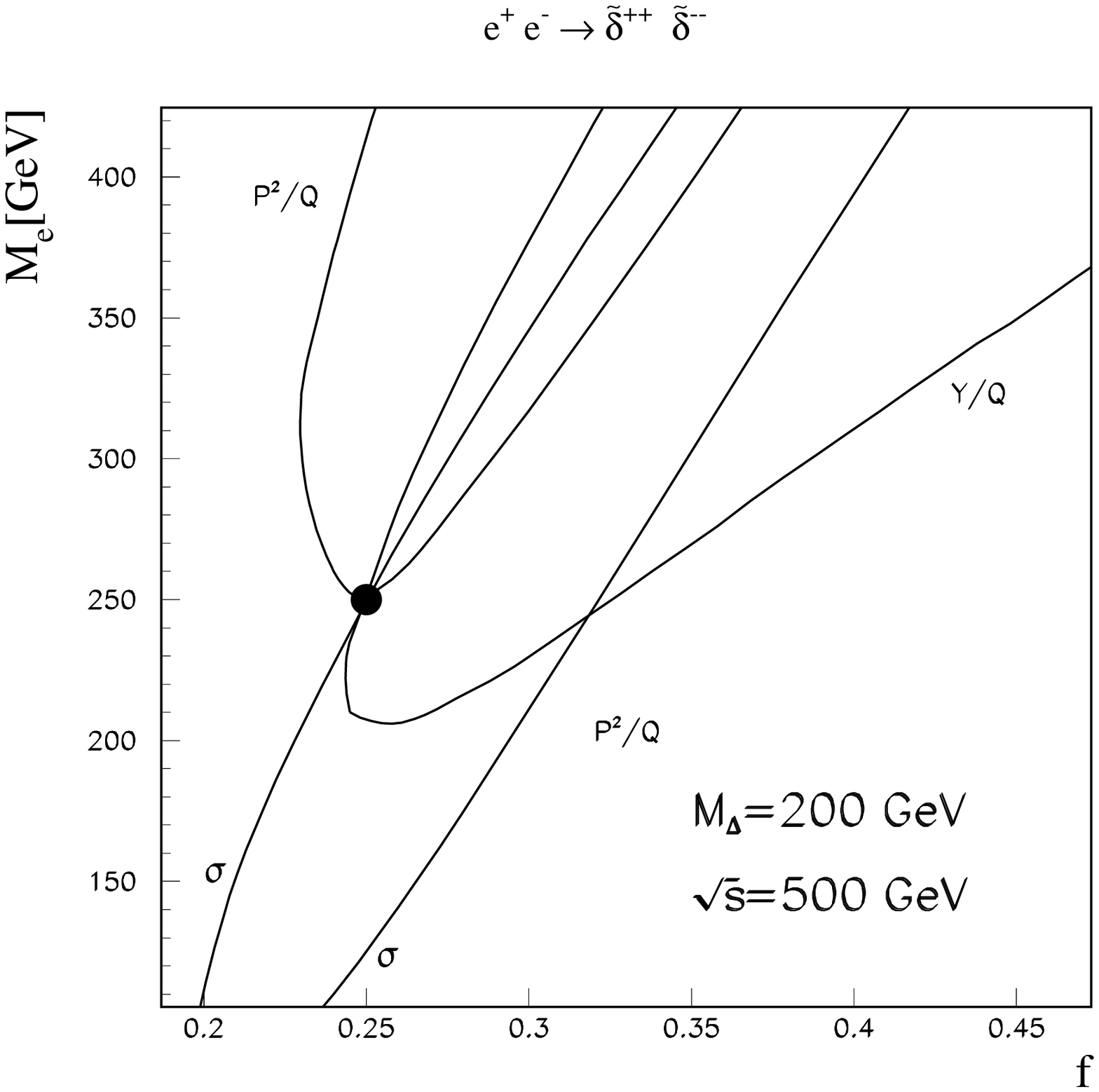}
}
%\hbox to\textwidth{\hss\epsfig{file=dpl.ps,height=7cm}\hss}
%\vskip -1.cm
%\end{center}
\caption{\it Contour lines from the measurements of 
 $\sigma,$  ${\cal P}^2/{\cal Q}$ and ${\cal Y}/{\cal Q}$
in the processes $\ee\ra\Dep\Dem$ and $\ee\ra\dep\dem.$
The common crossing point is indicated by the dot.
}
\label{fig:pqycurv}
\end{figure}

In the case of sufficiently light doubly charged higgsinos
and relatively heavy selectrons, the
cross sections and spin correlations can be exploited to determine
the selectron masses. If the selectrons are too heavy to be pair produced, 
their masses can be estimated from their virtual contributions  
to the higgsino  production processes.
This is demonstrated for the  ``measured values'' given above
by the [$ f, M_{\tilde e}$] contours in Figs.\ref{fig:pqycurv}.
In these examples, the isospin quantum numbers are
assumed to be pre--fixed in polarized--beam measurements.

%%%%%%%%%%%%%%%%%%%%%%%%%%%%%%%%%%%%%%%%%%%%%%%%%%%%%%%%
\subsubsection*{Acknowledgments}
%%%%%%%%%%%%%%%%%%%%%%%%%%%%%%%%%%%%%%%%%%%%%%%%%%%%%%%%

We are grateful to S. Ambrosanio,  S.-Y. Choi, U. Sarkar, G. Senjanovi{\'c},
M. Spira and F. Vissani for helpful discussions and K. Huitu
for useful comments.
M.R. thanks the Humboldt Foundation for a grant and  Prof. A. Wagner
for the warm hospitality extended to him at DESY.

\newpage

\end{document}